\documentclass[journal]{IEEEtran}
\usepackage{cite}
\usepackage{graphicx}
\usepackage{subcaption}
\usepackage{textcomp}
\usepackage[hidelinks]{hyperref}
\usepackage{amsmath}
\usepackage{algorithm}
\usepackage{array}
\usepackage{multirow}
\usepackage{tabulary}
\usepackage{courier}
\usepackage{booktabs}
\usepackage{amsmath}
\usepackage{amssymb}
\usepackage{algpseudocode}
\usepackage{float}
\usepackage{stfloats}
\usepackage{bm}
\usepackage{makecell}
\usepackage{xcolor}
\begin{document}

\title {Privacy-Preserving Cyberattack Detection in Blockchain-Based IoT Systems Using AI and Homomorphic Encryption}


\makeatletter
\newcommand{\linebreakand}{%
\end{@IEEEauthorhalign}
\hfill\mbox{}\par
\mbox{}\hfill\begin{@IEEEauthorhalign}
}
\makeatother
	
% \author{
%     \IEEEauthorblockN{Bui Duc Manh\IEEEauthorrefmark{1}, Chi-Hieu Nguyen\IEEEauthorrefmark{1}, Dinh Thai Hoang\IEEEauthorrefmark{1}, Diep N. Nguyen\IEEEauthorrefmark{1}, Ming Zeng\IEEEauthorrefmark{2}, and Quoc-Viet Pham\IEEEauthorrefmark{3}} \\
%     \IEEEauthorblockA{\IEEEauthorrefmark{1}School of Electrical and Data Engineering, University of Technology Sydney, Australia} \\
%     \IEEEauthorblockA{\IEEEauthorrefmark{2}Department of Electrical and Computer Engineering, Laval University, Quebec, Canada} \\
%     \IEEEauthorblockA{\IEEEauthorrefmark{3}School of Computer Science and Statistics, Trinity College Dublin, Ireland}
% }

\author{Bui Duc Manh, Chi-Hieu Nguyen, Dinh Thai Hoang, Diep N. Nguyen, Ming Zeng, and Quoc-Viet Pham
\thanks{Bui Duc Manh (corresponding author), Chi-Hieu Nguyen, Dinh Thai Hoang, and Diep N. Nguyen are with the School of Electrical and Data Engineering, University of Technology Sydney, Sydney, NSW 2007, Australia (e-mail: duc.m.bui-1@student.uts.edu.au; hieu.c.nguyen@student.uts.edu.au; hoang.dinh@uts.edu.au; diep.nguyen@uts.edu.au).}
\thanks{Ming Zeng is with the Department of Electrical Engineering and
Computer Engineering, Université Laval, Quebec, QC G1V 0A6, Canada (e-mail: ming.zeng@gel.ulaval.ca).}
\thanks{Quoc-Viet Pham is with the School of Computer Science and Statistics, Trinity College Dublin, The University of Dublin, D02 PN40, Ireland (e-mail: viet.pham@tcd.ie).} 
}
\maketitle
\begin{abstract}

This work proposes a novel privacy-preserving cyberattack detection framework for blockchain-based Internet-of-Things (IoT) systems. In our approach, artificial intelligence (AI)-driven detection modules are strategically deployed at blockchain nodes to identify real-time attacks, ensuring high accuracy and minimal delay. To achieve this efficiency, the model training is conducted by a cloud service provider (CSP). Accordingly, blockchain nodes send their data to the CSP for training, but to safeguard privacy, the data is encrypted using homomorphic encryption (HE) before transmission. This encryption method allows the CSP to perform computations directly on encrypted data without the need for decryption, preserving data privacy throughout the learning process. To handle the substantial volume of encrypted data, we introduce an innovative packing algorithm in a Single-Instruction-Multiple-Data (SIMD) manner, enabling efficient training on HE-encrypted data. Building on this, we develop a novel deep neural network training algorithm optimized for encrypted data. We further propose a privacy-preserving distributed learning approach based on the FedAvg algorithm, which parallelizes the training across multiple workers, significantly improving computation time. Upon completion, the CSP distributes the trained model to the blockchain nodes, enabling them to perform real-time, privacy-preserved detection. 
Our simulation results demonstrate that our proposed method can not only mitigate the training time but also achieve detection accuracy that is approximately identical to the approach without encryption, with a gap of around 0.01\%. Additionally, our real implementations on various blockchain consensus algorithms and hardware configurations show that our proposed framework can also be effectively adapted to real-world systems.
\end{abstract}


% \fi
\IEEEpeerreviewmaketitle

\section{Introduction}

\subsection{Motivation and Related Works}
\label{sec: related}

%\IEEEPARstart{I}{nternet}
% \color{blue}
The Internet of Things (IoT) has rapidly emerged as a transformative technology, powering intelligent applications across various domains, including smart homes, agriculture, and cities~\cite{iot2015Al}. With the proliferation of these applications, the number of IoT devices deployed worldwide has surged to approximately 15 billion in 2023, with projections suggesting this figure could double by 2030~\cite{sta2024}.
In IoT systems, the physical sensors collect data and send it to IoT gateways, which are responsible for monitoring and transmitting sensor data to the cloud server for further processing. Although the existing IoT infrastructure can provide effective automation and data management, it still faces various weaknesses that lead to the Single Point of Failure (SPoF) problem, remaining susceptible to cyberattacks, including Byzantine routing information attacks and injection attacks~\cite{iotsec2020},\cite{iot2019comst}.
% By relying on the high-end server provided by third-party services, the confidential data of IoT users can be read and tampered with by the cloud service provider, thereby leading to privacy concerns~\cite{iotsec2020}. 
% Moreover, data stored in the cloud remains susceptible to cyberattacks, including Byzantine routing information attacks and injection attacks~\cite{iotsec2020},\cite{iot2019comst}. 
To address these challenges, both industry and academia are increasingly adopting blockchain technology, widely acknowledged as a promising solution to significantly enhance the security and resilience of IoT architectures~\cite{iot2019comst}.

Blockchain is emerging as a potential technology that can significantly enhances the security, transparency, and reliability of IoT systems. Its decentralized ledger ensures that data is immutable and resistant to unauthorized manipulation or cyberattacks~\cite{iot2019comst},\cite{nakatomo2008}. In addition, blockchain enables secure, tamper-proof transactions between IoT devices, ensuring data integrity and trust in applications like smart homes, supply chain management, and industrial IoT~\cite{iot2019comst}. By removing the reliance on central authorities, it does not only prevents single points of failure but also improves system resilience~\cite{iot2019comst}.
Due to the above benefits, various industries have been actively leveraging blockchain to securely maintain IoT applications, such as IBM blockchain for supply chain applications and IOTA blockchain for healthcare data management~\cite{ibmsolution},\cite{iota}. 

Although the integration of blockchain with IoT fosters a more secure and efficient environment for the widespread adoption of IoT technologies, it is still vulnerable to multiple cyber threats. Statistics show that from 2011 to 2023, blockchain ecosystems have endured over 1,600 cyberattacks, including network attacks that caused substantial damage and contributed to financial losses exceeding \$32 billion~\cite{slowmist}.
% including significant network attacks, resulting in financial losses totally exceeding \$32 billion~\cite{slowmist}. 
% For instance, in August 2021, Poly Network, a cross-chain protocol for blockchain applications, reported that their system had been hacked, causing over \$611 million to be stolen~\cite{techmonitor}. Additionally, in March 2022, Ronin Network, an EVM-based blockchain game application, revealed that a hacker had successfully stolen multiple private keys, resulting in a loss of \$614 million~\cite{techmonitor}. 
% In terms of network security, there are also various network attacks on the blockchain environments. 
For instance, in April 2021, the Hotbit wallet, a blockchain cryptocurrency wallet, reported that their database was deleted by network attacks from hackers~\cite {slowmist}. Additionally, in September 2021, the Solana chain also reported that the system faced a network outage due to distributed denial of service (DDoS) attacks, leading to 12 hours offline of the chain~\cite{slowmist}. Most recently, in June 2024, BtcTurk, a Turkish cryptocurrency exchange, suffered a network attack that impacted ten wallets containing various cryptocurrencies, resulting in a \$5.3 million freeze~\cite{slowmist},~\cite{cointele}. These problems reveal persistent vulnerabilities within blockchain systems that could cause serious concerns about the security of blockchain-based IoT networks.

% To protect Blockchain-based IoT systems from cyberattacks, one can rely on intrusion detection methods to detect and prevent specific blockchain network attacks~\cite{fot2021},~\cite{wang2018eth}. The authors in~\cite{fot2021} analyzed the flooding of transactions (FoT) attack on the Moreno blockchain in terms of the network capacity, block size, etc., to evaluate the attacker's benefits. In~\cite{wang2018eth}, besides analyzing the cyberattacks (i.e., DDoS, brute passwords) on the Ethereum nodes, the authors also implemented the detection of cyberattacks based on scanning techniques. However, the above approaches can only detect several specific types of attacks and often detect threats only after significant damage has already been incurred.

% To protect blockchain-based IoT systems from cyberattacks, one can rely on intrusion detection methods to detect and prevent specific blockchain network attacks. These methods analyze attacks, such as DDoS or brute-force passwords, to design effective detection strategies~\cite{wang2018eth}. However, they can only detect several specific types of attacks and often detect threats only after significant damage has already been incurred.
% Recently, machine learning (ML) has emerged as an effective approach to developing intelligent and adaptive intrusion detection systems (IDS), improving performance by learning patterns from both normal and attack data~\cite{iotsec2020},~\cite{Hassan2023blockchain}.
To protect blockchain-based IoT systems from cyberattacks, one can rely on ML to effectively develop intelligent and adaptive intrusion detection systems (IDS)~\cite{iotsec2020},~\cite{Hassan2023blockchain}. By learning patterns from both normal and attack data, ML models can detect various attack types, including previously unknown threats, making them well-suited for blockchain environments that frequently encounter novel cyber threats~\cite{Hassan2023blockchain}. 
% \color{black}
In~\cite{kim2021bc}, the authors analyzed the Bitcoin traffic data and experimented with DoS and Eclipse attacks to collect network attack data. The authors then designed an autoencoder (AE) to detect the considered attacks on the Bitcoin traffic, which achieved high accuracy in detecting attacks. Moreover, the authors in~\cite{khoa2024bnat} designed a private Ethereum network to process data transmission from IoT devices. The benign and attack traffic collected from IoT devices and Ethereum nodes was used to train a Deep Belief Network (DBN), achieving impressive accuracy in detecting cyberattacks.

% Unlike the conventional methods, ML enhances the detection performance by allowing the model to learn from the distribution patterns of both normal and attack data, enabling it to identify multiple types of attacks~\cite{iotsec2020}. In addition, the ML model is capable of detecting new, previously unreported attacks, making it particularly effective for blockchain environments that frequently encounter novel cyber threats~\cite{Hassan2023blockchain}. In~\cite{kim2021bc}, the authors analyzed the Bitcoin traffic data and experimented with DoS and Eclipse attacks to collect network attack data. The authors then designed an autoencoder (AE) to detect the considered attacks on the Bitcoin traffic in which the accuracy of the considered AE reached nearly 99\%. Besides, the work in~\cite{Cao2021bc} utilized the Recurrent Neural Network (RNN) to detect various DDoS attacks on Ethereum networks. By simulating the Ethereum network and capturing both normal and attack behaviours, the approach demonstrated a 99\% accuracy in detecting attacks. Moreover, the authors in~\cite{khoa2024bnat} designed a private Ethereum network to process data transmission from IoT devices. The benign and attack traffic collected from IoT devices and Ethereum nodes was used to train a Deep Belief Network (DBN), achieving 98\% accuracy in detecting cyberattacks.

Nevertheless, in blockchain-based IoT systems, applying ML to detect cyberattacks raises substantial privacy concerns for users~\cite{privacyids2021},~\cite{acmsurvey2021ppml}.
In practice, ML-based systems typically rely on third-party services, such as cloud providers, to access the extensive computational resources required for high-demand tasks like training and inference~\cite{acmsurvey2021ppml}. This reliance necessitates data transfer to these third-party services, leading to potential privacy risks. While analyzing attack data, users' sensitive data, including personal details, biometric data, and healthcare records, could be extracted and accessed by these third-party providers, compromising data confidentiality~\cite{Hui2021privacy}. Although various distributed techniques like federated learning and collaborative learning have been developed to reduce dependency on centralized training services~\cite{acmsurvey2021ppml, hoang2020fl}, they require data owners to train their own data before sharing their trained models to improve overall detection performance. This approach is limited, as many data owners, such as IoT gateways, often lack the hardware capabilities to fully train their data. 
To overcome the considered privacy challenge of centralized ML aproaches, Homomorphic Encryption (HE) offers a compelling solution for integration with ML models. By employing this lattice-based cryptographic method, data can be encrypted before being sent to third-party services, enabling ML models to perform computations directly on the encrypted data without the need for decryption~\cite{miran2018logistic}. This ensures that user privacy is preserved throughout the entire process. As a result, HE stands out as a highly effective enhancement for ML models, providing robust protection for users' privacy.

To protect data privacy, several studies have successfully integrated HE into ML models. For instance, the authors in~\cite{doren2021} introduced Doren, a deep learning-based HE method for processing of Convolutional Neural Networks (CNNs) on encrypted data. By employing the Single Instruction Multiple Data (SIMD) packing and bootstrapping techniques, they enabled HE computation on prominent CNN models, such as VGG7 and ResNet20, achieving accuracy from 73.85\% to 92.32\% within multiple HE schemes and DL models. Similarly, in~\cite{miran2022secure}, the authors proposed a CNN-integrated HE method to recognize human actions in a privacy-preserved manner. By utilizing the SIMD packing method for efficient encrypted matrix multiplication, their approach results in a high inference throughput of 0.4 to 0.8 seconds per sample and accuracy from 86\% to nearly 89\% within multiple datasets and CNN configurations. However, it is worth noting that while these studies successfully handle the inference process in ML-integrated HE, they do not address the complexities of the training process, which is arguably the most critical phase in machine learning.

% To preserve the privacy of users, various works have integrated HE into ML models.
% In~\cite{cryptonets2016}, the authors proposed a neural network-integrated with HE utilizing non-linear activation functions. By evaluating the considered neural network on the MNIST dataset, the authors showed that the HE-encrypted neural network can achieve the accuracy of 99\% and high throughput in the encryption and decryption process. 
% Additionally, the authors in~\cite{doren2021} introduced Doren, a DL-based HE method that allows the processing of a Deep Convolutional Neural Network (CNN) over HE-encrypted data. By employing the SIMD packing and bootstrapping techniques, the authors enable the HE computation on multiple famous CNN models (i.e., VGG7, ResNet20). The simulation results indicated that the proposed approach could achieve accuracy from 73.85\% to 92.32\% within multiple HE schemes and DL models. Besides, in~\cite{miran2022secure}, the authors proposed a CNN-integrated HE method to recognize human activities in a privacy-preserved manner. The authors also based on the SIMD packing method to design effective HE-encrypted matrix multiplication, forming an effective encrypted CNN with high throughput during inference. The results showed that the proposed approach could have a high throughput of 0.4 to 0.8 seconds per sample and accuracy from 86\% to nearly 89\% within multiple datasets and CNN configurations. However, these works only focus on the inference process of ML-integrated HE without considering the training process. 

In~\cite{hesamifard2018privacy}, the authors proposed CryptoDL, which considered both training and inference on the HE-encrypted data. However, the proposed CryptoDL requires frequent communication between ML owners and users to refresh the encrypted parameters, preventing the overload of HE noise during the training process. Alternatively, the proposed approach costs massive training time, resulting in around 1,456.7 seconds to train over one iteration with a 5-layer neural network. The authors in~\cite{Nandakumar_2019_CVPR_Workshops} also considered the training task of neural network-integrated HE with a multi-threading approach. The results showed that the training task of a 3-layer neural network requires intensive training time. Specifically, it takes over 9 hours with a single thread and 40 minutes with 30 threads to process a mini-batch of 60 samples. In~\cite{hieu2024enc}, the authors developed a training algorithm for an encrypted neural network that achieves nearly 88\% accuracy in recognizing human activities, yet they did not take the training time into consideration. 

\subsection{Challenges and Contributions}
\label{sec:challenge}

As discussed above, relying on ML to enhance blockchain-based IoT security raises notable privacy concerns. Given HE's proven effectiveness in safeguarding privacy when combined with ML, training DL models on HE data presents two significant challenges, i.e., computational overhead and limited efficient operations. Generally, HE enables computations on encrypted data without needing to decrypt it, which is crucial for privacy-preserving ML~\cite{acmsurvey2021ppml}. However, it comes at the cost of extensive computational overhead, as operations on ciphertext require much more resources than on non-encrypted data. Therefore, the extensive computational requirement for training tasks of ML-integrated HE leads to prohibitively long training time, making it challenging to scale DL models to large datasets when preserving users' privacy by HE. Moreover, \textbf{since HE schemes only efficiently support a limited set of mathematical operations (i.e., element-wise addition and multiplication)}, effectively implementing ML training tasks for HE-encrypted data (matrix multiplication, back-propagation, non-linear functions, etc.) is particularly challenging.

To address all the above challenges, we propose a novel privacy-preserving cyberattack detection framework for blockchain-based IoT systems. In the proposed system, AI-based smart cyberattack detection modules are deployed at the mining nodes in the blockchain network to detect attacks on the mining nodes in a real-time manner. To improve the efficiency in detecting attacks in real-time (i.e., with high accuracy and low delay), the training process is performed in advance at a cloud service provider (CSP). In particular, the mining nodes send their training data to the CSP for a comprehensive training process. To protect data privacy, before sending the training data to the CSP, the mining nodes encrypt their data using the HE technique. This technique allows the CSP to perform global model training on encrypted data without the need to decrypt it, thereby protecting data privacy. To address the problem of handling a huge amount of encrypted data at the CSP, we first develop an innovative packing algorithm to pack the data in an SIMD manner, thereby effectively enabling the training process for HE-encrypted data. After that, we design an innovative training algorithm for the deep neural network with HE-encrypted data based on our proposed packing methods. 
While applying HE to ML/DL models can effectively enhance data privacy, it presents significant challenges for the CSP regarding computation time during the training task. To tackle this, we propose a privacy-preserving distributed learning for HE-encrypted data. Our proposed approach allows the learning model to be trained in parallel across multiple workers, thereby improving computation time. Once the training process is completed, the CSP will share the trained model with the mining nodes for real-time detection. Our main contributions are summarized as follows:
\begin{itemize}
    % \item We introduce a novel privacy-preserving cyberattack detection framework in IoT-based blockchain networks. By leveraging the cloud's resources and HE, the system can effectively detect cyberattacks in a decentralized environment while securely maintaining the privacy of users.
    \item We develop innovative packing methods for one-dimensional (1D) vectors and two-dimensional (2D) matrices, which allow the encrypted matrix multiplication in an SIMD manner. Following that, we design a robust training algorithm for encrypted data based on the proposed packing algorithm mentioned above.
    \item We propose a privacy-preserving distributed learning algorithm that significantly optimizes the training time over the encrypted data. By leveraging the FedAvg algorithm, computing resources from multiple parties can be utilized to form an effective learning approach. The experiments show that the proposed learning algorithm reduces training time significantly as more workers participate.
    \item We conduct comprehensive evaluations on the detection performance with the real-world blockchain network attack dataset. The results show that our approach not only delivers highly accurate cyberattack detection for both encrypted and non-encrypted inference but also achieves consistency in performance comparable to the non-encrypted baseline method. 
    
    % This validates the robustness and reliability of our solution in maintaining privacy without compromising detection accuracy.
    \item We perform real experiments to evaluate our proposed framework across various consensus mechanisms and hardware configurations. The results indicate that our framework is highly adaptable to practical applications, delivering impressive efficiency in resource utilization, low latency, and high throughput. 
    
    % These findings underscore the framework's potential for seamless integration into real-world systems, meeting the demands of modern decentralized environments. 
\end{itemize}

The rest of the paper is organized as follows. In Section~\ref{sec:sysmodel}, we introduce the fundamental background of the blockchain-based IoT system and describe our proposed privacy-preserving cyberattack detection framework. In Section~\ref{sec:proposed_ppdl}, we first present the fundamentals of HE and then introduce the proposed training algorithm for HE-encrypted deep neural network and privacy-preserving distributed learning algorithm. Section~\ref{sec: performance_eval} provides comprehensive evaluations of our proposed framework with a real-world blockchain network attack dataset. Finally, we conclude our paper in Section~\ref{sec: conclu}.

% \begin{figure}[t]
%     \centering
%     \includegraphics[width=\linewidth]{Figures/System_Model_main.pdf}
%     \caption{The proposed privacy-preserving distributed learning framework for cyberattack detection.}
%     \label{fig:sys}
% \end{figure}

\begin{figure*}[t]
    \centering
    \includegraphics[width=\textwidth]{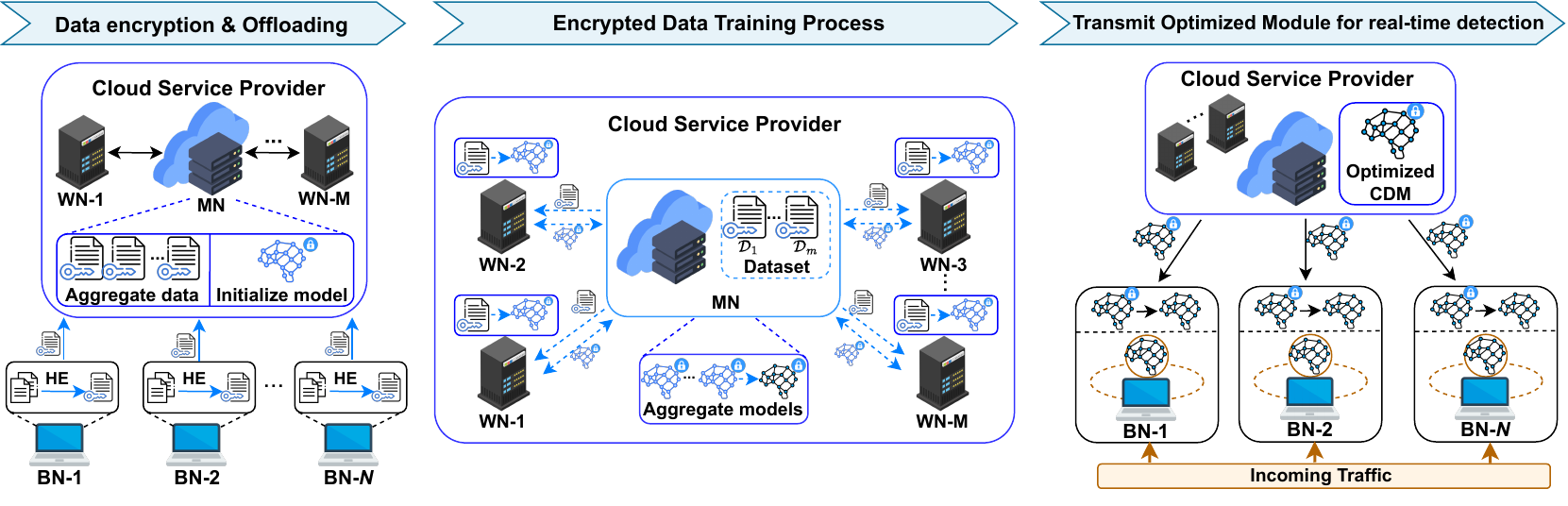}
    \hfill
    \includegraphics[width=\linewidth]{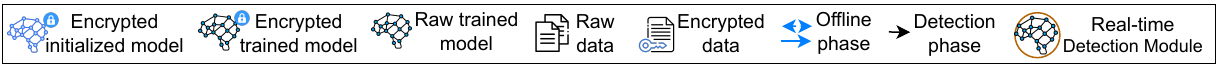}
    \caption{The proposed privacy-preserving cyberattack detection framework includes three phases: data encryption and offloading, encrypted data training, and real-time detection. The CSP operates as a network security service to ensure the security of $N$ blockchain nodes (BNs).}
    \label{fig:sys}
\end{figure*}

% \begin{figure*}[t]
%     \centering
%     \begin{subfigure}[b]{\textwidth}
%         \centering
%         \includegraphics[width=\linewidth]{Figures/System_Model.pdf}
%         \caption{Offline training phase}
%         \label{fig:sys1}
%     \end{subfigure}%
%     \hfill
%     % \begin{subfigure}[b]{0.5\textwidth}
%     %     \centering
%     %     \includegraphics[width=\linewidth]{Figures/Sys_model2.pdf}
%     %     \caption{Cyberattack Detection phase}
%     %     \label{fig:sys2}
%     % \end{subfigure}
%     % \hfill
%     \begin{subfigure}{\linewidth}
%         \centering
%         \includegraphics[width = \linewidth]{Figures/Image_labels.pdf}
%     \end{subfigure}
%     \caption{The proposed privacy-preserving distributed cloud-native learning framework for cyberattack detection includes (a) offline training phase and (b) cyberattack detection phase. The CSP operates as a network security service to ensure the security of $N$ blockchain nodes, including light nodes and full nodes.}
%     \label{fig:sys}
% \end{figure*}

\section{System Model and Proposed Cyberattack Detection Framework}
\label{sec:sysmodel}

% Before describing our privacy-preserving cyberattack detection framework, we first provide a general overview of blockchain-based IoT systems.
% \begin{figure}[t]
%     \centering
%     \begin{subfigure}[b]{0.4\textwidth}
%         \centering
%         \includegraphics[width=\linewidth]{Figures/System_Model.pdf}
%         % \captionsetup{justification=raggedright, singlelinecheck=false, margin=2cm}
%         \caption{Pre-learning process}
%         \label{fig:pre_model}
%     \end{subfigure}%
%     \hfill
%     \begin{subfigure}[b]{0.4\textwidth}
%         \centering
%         \includegraphics[width=\linewidth]{Figures/System_Model2.pdf}
%         % \captionsetup{justification=raggedright, singlelinecheck=false, margin=2cm}
%         \caption{Privacy-preserving learning process}
%         \label{fig:main_model}
%     \end{subfigure}
%     \caption{The proposed privacy-preserving intrusion detection framework including pre-learning and privacy-preserving learning.}
%     \label{fig:sys}
% \end{figure}

\subsection{Overview of Blockchain-based IoT System}
\label{sec:iot-bc}
% Blockchain is an innovative digital distributed ledger that provides a decentralized, transparent and secure way of storing and managing data across multiple parties. By utilizing the decentralized feature of peer-to-peer networks, each participant can maintain a copy of the ledger, therefore eliminating the need for centralized authority.
% However, when integrating blockchain to IoT systems, many devices are constrained by limited resources, which makes it challenging for them to hold a copy of the ledger or engage in the mining process, such as validating or publishing new blocks~\cite{iot2019comst}. These resource limitations necessitate tailored approaches for integrating IoT devices (e.g., gateways and sensors) into blockchain networks, ensuring that they can participate effectively without being overburdened. In such scenarios, IoT devices may focus on transmitting data to more powerful nodes in the network (e.g. blockchain mining nodes), which handle the extensive computational tasks associated with blockchain operations~\cite{iot2019comst}. This approach allows IoT systems to function based on blockchain, harnessing its decentralization, security, and transparency in IoT data management while maintaining efficiency and scalability despite the constraints of individual devices.
% \color{blue}
Blockchain offers a peer-to-peer and secure framework for IoT systems, enhancing decentralization, data integrity and trust~\cite{iot2019comst}. While constrained IoT devices may struggle with resource-intensive tasks like maintaining a full ledger or mining, IoT gateways can bridge this gap by handling key blockchain functions. These gateways enable the integration of blockchain into IoT systems, ensuring security, transparency, and efficient data management without overburdening individual devices.
There are several types of blockchain networks, each tailored to meet the specific needs of different IoT applications, including Public Blockchain, Private Blockchain, and Consortium Blockchain~\cite{iot2019comst}. The public blockchain permits any user to participate in the network, in which the users are required to pay an incentive fee to send transactions. In contrast, private and consortium blockchains restrict access to only authorized nodes while removing the need for transaction fees, enabling the deployment of applications for confidential purposes. Although the type of employed blockchain depends on the purposes of IoT applications, blockchain-based IoT systems still operate using the same approach. In general, the IoT devices send the transactions consisting of IoT data to the blockchain nodes for the mining process. Based on different consensus mechanisms (e.g. PoW, PoS, PBFT), the validators (miners) will store transactions into a block and compete to become the block leaders~\cite{cong2019pos}. Once the new block is verified on the chain, it is linked to the previous block via the hash value of the block's header, making it nearly impossible to alter the recorded information in the block and thereby ensuring the security and immutability of the blockchain~\cite{nakatomo2008}. In summary, the ability of IoT devices to participate in the blockchain enables IoT systems to achieve a more secure, immutable, and decentralized network.
% \color{black}

\subsection{Privacy-Preserving Cyberattack Detection in Blockchain Networks}
\label{sec:pp_sys}

The proposed privacy-preserving cyberattack detection system is illustrated in Fig.~\ref{fig:sys}. The system consists of a CSP consisting of a master node (MN) along with $M$ worker nodes (WNs), and $N$ blockchain nodes (BNs). 
In our framework, we consider an IoT network operated by an IoT Service Provider (IoTSP). Similar to~\cite{ibm}, the IoTSP deploys a private blockchain network to manage the IoT service and transactions, thereby maintaining trust between blockchain nodes and IoT endpoints. In this regard, cyberattacks on blockchain nodes can potentially take control of the transactions in blockchain networks, and thus Cyberattack Detection Modules (CDMs) can be deployed at the blockchain nodes to detect and prevent cyberattacks. Nevertheless, deploying the CDMs on the blockchain nodes for real-time detection is challenging. As mentioned in Section~\ref{sec:iot-bc}, since the blockchain nodes must perform various functions to maintain the blockchain (e.g., mining, validating, verifying, and storing blocks), training the CDMs on these nodes is inefficient due to extensive computational demand of training task. Although the CSP with extensive cloud computing resources is an effective solution to support the IoTSP within the training and deployment of the CDM, processing the CDMs externally on a cloud server raises significant concerns regarding user data privacy. To address this, our proposed framework employs the HE to protect user privacy while ensuring accurate deployment of CDMs on blockchain nodes. Specifically, our privacy-preserving cyberattack detection framework includes three phases: data offloading phase, encrypted data training phase, and real-time detection phase, as illustrated in Fig.~\ref{fig:sys}.

% \color{blue}
\textbf{Data encryption and offloading:} In the data encryption and offloading phase, the local dataset of each BN is offloaded to the CSP for further processing, as shown in Fig.~\ref{fig:sys}. However, before transmitting data to the CSP, BNs generate HE key pairs and use them to encrypt the uploading data, in which the HE scheme is introduced in Section~\ref{sec:he}. Once the CSP receives the encrypted data from blockchain nodes, it assembles it into a large encrypted dataset and then initializes a learning model specifically designed for cyberattack detection to prepare for the next phase. 

\textbf{Encrypted data training process:} The CSP starts the training phase by initially evaluating the number of available WNs within the cloud environment. Based on this assessment, the MN strategically partitions the full encrypted dataset into multiple segments and then distributes each across the predefined WNs. In this way, the computing resources of the active WNs are leveraged to maximize the training efficiency of the large encrypted dataset, reducing the computational overhead challenge of HE-encrypted data training. Upon receiving the corresponding encrypted partition, each WN uses the initialized learning model to train its assigned segment using our proposed training algorithm for encrypted data, detailed in Section~\ref{sec: training_ago}. After completing each learning round, the WNs send the encrypted trained models to the MN, which aggregates them into an updated encrypted model (the aggregation of the encrypted models is detailed in Section~\ref{sec:implementation_ppdl}). This updated model is then transmitted back to WNs to begin the next learning round. The CSP maintains this iterative process until the considered learning model converges or the predetermined number of learning rounds is reached, ensuring the model is fully optimized for the next phase. 

\textbf{Real-time detection:} During the real-time cyberattack detection phase, the CSP sends the optimized encrypted model back to the BNs for local deployment. Once the BNs receive the model, they decrypt the model to form a real-time detection module which can detect and prevent incoming attack traffic to the blockchain network.

\section{The Privacy-Preserving Distributed Learning}
\label{sec:proposed_ppdl}

As we described in Section~\ref{sec:sysmodel}, the CSP is responsible for training the learning model-based HE with a large encrypted dataset collected from blockchain nodes. However, as discussed in Section~\ref{sec:challenge}, training deep learning models on HE-encrypted data presents two major challenges: the lack of optimized training algorithms specifically designed for encrypted data and the significant computational overhead that results in extensive training time.
To address these problems, we initially propose an effective training algorithm for the deep neural network with HE-encrypted data. Following that, we design the distributed training approach to optimize the processing time of our training algorithm. Before detailing our proposed algorithm, we first introduce the fundamental concepts of HE.

\subsection{Homomorphic Encryption}
\label{sec:he}

Homomorphic encryption is a cryptography technique allowing mathematical operation over encrypted data without decryption~\cite{gentry2009fully}. In order to integrate HE with the neural network, we propose to use the Cheon-Kim-Kim-Song (CKKS) scheme, which enables approximate calculations on floating-point numbers~\cite{cheon2017ckks}. 
% Additionally, the CKKS allows for the encryption of multiple plaintexts within a single ciphertext, facilitating parallel computation on encrypted vector in a SIMD manner~\cite{cheon2017ckks}. 
Additionally, the CKKS enables the encoding of multiple data segments into a single plaintext, which is subsequently encrypted into a ciphertext, facilitating parallel computation on encrypted vector in a SIMD manner~\cite{cheon2017ckks}. 
Since the CKKS scheme is based on the ring learning with errors (RLWE), it requires the ring dimension $\mathcal{R}$ (a power-of-two integer) to ensure the security level and multiplicative depth~\cite{gentry2009fully}. Following that, the size of ciphertext (i.e., the maximum number of data segments it can contain) is denoted as $\mathcal{B}$, in which $\mathcal{B} = \mathcal{R}/2$. It is worth noting that the size of a plaintext is equal to that of its corresponding ciphertext. To be more specific, the CKKS scheme includes the key generation, encryption, decryption, and basic homomorphic operations as follows:

\begin{itemize}
    \item \textsf{SKGen}$(n)$: generate random secret key $\textsf{sk}_n$ for user $n$.
    \item $\textsf{PKGen}(\textsf{sk}_n)$: create the public key $\textsf{pk}_n$ for user $n$ based on the secret key $\textsf{sk}_n$.
    \item $\textsf{Enc}(\textsf{pk}_n, c)$: encrypt a raw vector $\mathbf{c}$ into a ciphertext $\bm{\hat{c}}$ by using the public key $\textsf{pk}_n$.
    \item $\textsf{Dec}(\textsf{sk}_n, \bm{\hat{c}})$: decrypt encrypted vector $\bm{\hat{c}}$ into its plain form $c$ by using the secret key $\textsf{sk}_n$
    % \item $\textsf{EvalAdd}(\bm{\hat{c}_1},\bm{\hat{c}_2})$: 
    % \item $\textsf{EvalSub}(\bm{\hat{c}_1},\bm{\hat{c}_2})$:
    % \item $\textsf{EvalMul}(\bm{\hat{c}_1},\bm{\hat{c}_2})$:
    \item $\textsf{Add}(\bm{\hat{c}_1},\bm{\hat{c}_2})$: the addition between two ciphertexts $\bm{\hat{c}_1}$ and $\bm{\hat{c}_2}$, in which $\textsf{Dec}(\textsf{sk}_n, \textsf{Add}(\bm{\hat{c}_1},\bm{\hat{c}_2})) \approx \mathbf{c_1} + \mathbf{c_2}$.
    \item $\textsf{Sub}(\bm{\hat{c}_1},\bm{\hat{c}_2})$: the subtraction between two ciphertexts $\bm{\hat{c}_1}$ and $\bm{\hat{c}_2}$, in which $\textsf{Dec}(\textsf{sk}_n, \textsf{Sub}(\bm{\hat{c}_1},\bm{\hat{c}_2})) \approx \mathbf{c_1} - \mathbf{c_2}$.
    \item $\textsf{Mult}(\bm{\hat{c}_1},\bm{\hat{c}_2})$: the multiplication between two ciphertexts $\bm{\hat{c}_1}$ and $\bm{\hat{c}_2}$, in which $\textsf{Dec}(\textsf{sk}_n, \textsf{Mult}(\bm{\hat{c}_1},\bm{\hat{c}_2})) \approx \mathbf{c_1} \times \mathbf{c_2}$.
\end{itemize}

\begin{figure}[t]
    \centering
    \begin{subfigure}[b]{0.48\textwidth}
        \centering
        \includegraphics[width=\linewidth]{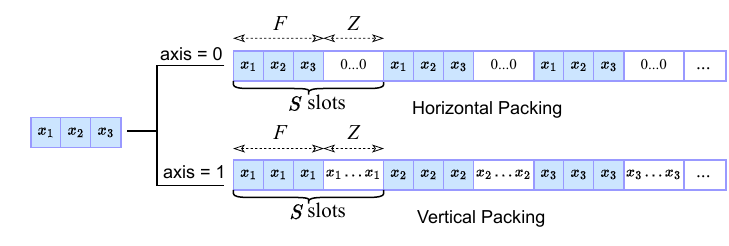}
        \caption{1D packing for a vector}
        \label{fig:pack1d}
    \end{subfigure}%
    \hfill
    \begin{subfigure}[b]{0.48\textwidth}
        \centering
        \includegraphics[width=\linewidth]{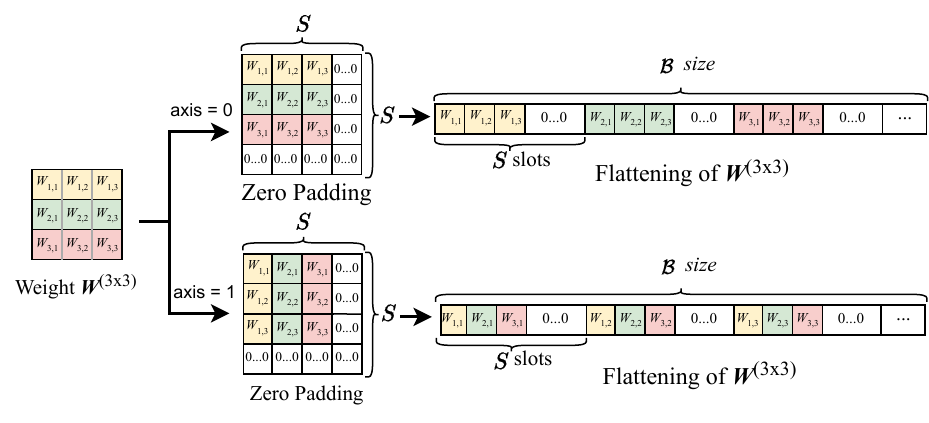}
        \caption{2D packing for a matrix}
        \label{fig:pack2d}
    \end{subfigure}
    \caption{Illustration of the proposed alternative packing method, including (a) 1D packing and (b) 2D packing. The proposed methods are the pre-processing of HE, which fits the elements of data (i.e., vector and matrix) into the slots of a ciphertext in both normal (axis=0) and transpose (axis=1) manners. Therefore, this enables efficient encrypted matrix/vector multiplications during the training task.}
    \label{fig:pack_algo}
\end{figure}

It is worth noting that the aforementioned homomorphic evaluations (i.e., addition, subtraction, and multiplication) perform element-wise operations between ciphertexts to produce the encrypted outputs. However, the more homomorphic evaluations on ciphertext, the more noises are added, which causes the error in decryption~\cite{gentry2009fully}. To solve that problem, the bootstrapping mechanism denoted as $\textsf{Bootstrap}(\bm{\hat{c}})$ can be applied to reduce the magnitude of noise in ciphertext $\bm{\hat{c}}$ by re-encrypting it~\cite{gentry2009fully}. This capability enables additional computation on ciphertext, making it suitable for deep learning tasks, including training and inference. The proposed training procedure for HE-encrypted data will be detailed as follows.

% In addition to the basic evaluation algorithms, we also utilize the \textsf{SumCols} and \textsf{SumRows} algorithms provided by OpenFHE open-source library~\cite{openfhe}, which are designed based on \textsf{AllSum} algorithm in~\cite{miran2018logistic}. To be more specific, we consider a ciphertext representing multiple plain vectors $(x_1,\dots, x_l)$ .The \textsf{SumCols} and \textsf{SumRows} can be denoted as follows:
% \begin{equation}
%     \textsf{SumCols}(\bm{\hat{c}}, S) = 
% \end{equation} 

\subsection{The Proposed Training Process of Deep Neural Network for HE-Encrypted Data}
\label{sec: training_ago}

\begin{algorithm}[t]
\caption{\textsf{Pack1D} for packing 1D vector}
\label{ago:pack1d}
\begin{algorithmic}[1]
\State \textbf{Input:} $x, S, \mathcal{B}, \bar{x}_{\text{axis}}$, with $x = [x_1,x_2,\dots, x_n]$ and $\bar{x}_{\text{axis}} \in \{0,1\}$
\State \textbf{Output:} $\bar{x}$
\State Initialize $step = \frac{\mathcal{B}}{S}$
\If{$\bar{x}_{\text{axis}}=0$}
    \State $x = \textsf{ZeroPad}(x,S)$
    \State \parbox[t]{0.8\linewidth}{$\bar{x} = \textsf{Replicate}(x,step)$ which $\bar{x}_{i+j\cdot n} = x_i$, $i \in \{1,2,\dots,n\}$ and $j \in \{0,1,\dots,step-1\}$}
\Else
    \State \parbox[t]{0.8\linewidth}{ $\bar{x} = \textsf{Repeat}(x,S)$ which $\bar{x}_{(i-1)\cdot S + j} = x_i$, $i \in \{1,2,\dots,n\}$ and $j \in \{1,2,\dots,S\}$}
\EndIf
\State $\bar{x} = \textsf{ZeroPad}(\bar{x},\mathcal{B})$
S\State \Return $\bar{x}$
\end{algorithmic}
\end{algorithm}

\subsubsection{The Proposed Packing Methods}

For efficient homomorphic evaluation during the training task, we first develop two alternative packing approaches called \textsf{Pack1D} and \textsf{Pack2D}. To be more specific, \textsf{Pack1D} is applied to process 1D data and then produce the output in corresponding $\bar{x}_{\text{axis}}$, in which $\bar{x}_{\text{axis}}$ is the predefined axis of the output (e.g., $\bar{x}_{\text{axis}} = 0$ for horizontal axis and $\bar{x}_{\text{axis}} = 1$ for vertical axis). It is worth noting that in HE, a plaintext contains multiple segments, with the number of slots $S$ in each segment, and each slot holds an individual value. As illustrated in Algorithm~\ref{ago:pack1d}, \textsf{Pack1D} takes the input consisting of a 1D array, slot size of a segment, size of ciphertext, and the considered axis of output, which are respectively defined as $x$, $S$, $\mathcal{B}$, and $\bar{x}_{\text{axis}}$. As described in Fig.~\ref{fig:pack1d}, the axis for encryption (i.e., horizontal or vertical) is first determined. Regarding the horizontal axis, the sample is zero-padded to match with a plaintext slot size $S$, where $S = \lfloor\sqrt\mathcal{B}\rfloor$ and $S = F + Z$, with $F$ being the number of values in the sample and $Z$ being the number of added zeros. Notably, $F$ is typically less than $S$ as network traffic data, the focus of this paper, often has a limited number of features. Then, it is replicated $\mathcal{B}/S$ times to fit with the size of the ciphertext. Otherwise, if the axis is vertical, each element of the sample is repeated $S$ times, forming a new array, which is then zero-padded to fit with the size of the ciphertext. 

The proposed \textsf{Pack2D} algorithm is described in Algorithm~\ref{ago:pack2d}, which is applied to pack a 2D matrix into a 1D vector. Similar to \textsf{Pack1D}, \textsf{Pack2D} initially determines the axis for encryption. If the axis is vertical, the matrix is transposed; otherwise, it remains unchanged. Subsequently, the matrix is zero-padded to a square matrix with the size of $S \times S$. This square matrix is then flattened by concatenating its rows to form a 1D vector. For more clarity, the implementation of our 2D packing method is shown in Fig.~\ref{fig:pack2d}. 

To perform encryption based on our proposed packing methods, we consider a dataset $\mathcal{D} = (\textbf{x}_i,\textbf{y}_i)$ consisting of $I$ samples where $i \in \{1,\dots, I\}$. Here, $\textbf{x}_i$ and $\textbf{y}_i$ are, respectively, the training samples and labels, in which $\textbf{x}_i$ is a feature vector of i-th sample attached with a one-hot encoded label $\textbf{y}_i$. 
Based on the aforementioned HE operations, the key pair, including the public key and secret key, are used to generate the encrypted dataset $\hat{\mathcal{D}}$. Before the encryption of the dataset, we apply the 1D packing algorithm, described in Algorithm~\ref{ago:pack1d} as the preprocessing step for each sample. Therefore, the encryption process of encrypted feature vectors and encrypted labels can be denoted as:
\begin{equation}
\label{eq:enc_x}
   \hat{\bm{x}}_i = \textsf{Enc}\big(\textsf{Pack1D}(\textbf{x}_i, \bar{x}_{\text{axis}}, S, \mathcal{B}), \textsf{pk}\big),
\end{equation} 
\begin{equation}
\label{eq:enc_y}
   \hat{\bm{y}}_i = \textsf{Enc}\big(\textsf{Pack1D}(\textbf{y}_i, \bar{y}_{\text{axis}}, S, \mathcal{B}), \textsf{pk}\big),
\end{equation} 

\noindent where $\hat{\bm{x}}_i$ and $\hat{\bm{y}}_i$ ($\forall i \in \{1,\dots, I\}$) are the encrypted feature vector and its attached encrypted label of the dataset $\hat{\mathcal{D}}$, respectively. This encrypted dataset will be the input for a deep neural network consisting of $K$ layers, in which its plain form can be represented as:
\begin{equation}
   h^{(k)} = \sigma(W^{(k)}\cdot h^{(k-1)} + b^{(k)}), \quad k \in \{1,\dots, K\},
\end{equation} 

\noindent where $W^{(k)}$, $h^{(k)}$, $b^{(k)}$, $\sigma$ are parameters of the neural network consisting of weights, the output of layer $k$, biases, and activation function, respectively. Here, $h^{(0)}$ is the input feature vector $x_{in}$ while $h^{(K)}$ is the output of the neural network $x_{out}$. In conventional deep learning, these parameters will be applied to produce the optimized output $x_{out}$ via the training process. Here, to process HE-encrypted data, the weights and biases are also encrypted to homomorphic form. Before performing encryption, the weights will be preprocessed by our proposed 2D packing algorithm, which is illustrated in Algorithm~\ref{ago:pack2d}. In particular, the encryption process is denoted as:
\begin{equation}
\label{eq:enc_weight}
   \hat{\bm{W}}^{(k)} = \textsf{Enc}\big(\textsf{Pack2D}(W^{(k)}, \bar{W}_{\text{axis}}^{(k)}, S, \mathcal{B}), \textsf{pk}\big),
\end{equation} 

\begin{algorithm}[t]
\caption{\textsf{Pack2D} for packing 2D matrix}
\label{ago:pack2d}
\begin{algorithmic}[1]
\State \textbf{Input:} $\textit{X}, S, \mathcal{B}, \bar{x}_{\text{axis}}$, with $X \in \mathbb{R}^{m \times n}$ and $\bar{x}_{\text{axis}} \in \{0,1\}$
\State \textbf{Output:} $\bar{x}$
\If{$\bar{x}_{\text{axis}}=1$}
    \State $\bar{X} = \textsf{Transpose}(X)$
\Else 
    \State $\bar{X} = X$
\EndIf
\State $\bar{X} = \textsf{ZeroPad}(\bar{X},S)$ which $m\times n = S^2$
\State $\bar{x} = \textsf{Flatten}(\bar{X})$
\State \Return $\bar{x}$
\end{algorithmic}
\end{algorithm}

\noindent where $\hat{\bm{W}}^{(k)}$ is the encrypted weight matrix of layer $k$.
To allow the multiplication of the weight matrix in encrypted form, the axis of $\hat{\bm{W}}^{(k)}$ is varied based on k-th layer. In particular, the value of $\bar{W}_{\text{axis}}$ can be defined as:
\begin{equation}
    \bar{W}_{\text{axis}}^{(k)} = 
    \begin{cases}
        0, & \text{if } k \textit{ mod } 2 = 1,\\
        1, & \text{if } k \textit{ mod } 2 = 0, 
    \end{cases}
\end{equation}

\noindent An axis value of 0 enables the computation along the row-based axis, whereas the axis value of 1 facilitates the column-based processing, as depicted in Fig.~\ref{fig:pack_algo}. Following that, the biases will be processed by applying the 1D packing algorithm before the encryption, which is denoted as:
\begin{equation}
\label{eq:enc_bias}
   \hat{\bm{b}}^{(k)} = \textsf{Enc}\big(\textsf{Pack1D}(b^{(k)}, \bar{b}_{\text{axis}}^{(k)}, S, \mathcal{B}), \textsf{pk}\big),
\end{equation} 
where $\hat{\bm{b}}^{(k)}$ is the encrypted bias vector of layer $k$. Thus, the output of the $\hat{\bm{W}}^{(k)}$ is the ciphertext on the orthogonal axis of its encrypted weight. For instance, the encrypted weight matrix packed on a row-based axis will generate output on a column-based axis. Hence, the axis of encrypted bias is calculated as: 
\begin{equation}
    \bar{b}_{\text{axis}}^{(k)} = 1 - \bar{W}_{\text{axis}}^{(k)}.
\end{equation}
By leveraging our proposed packing methods to process the encrypted input and encrypted parameters of the neural network in an alternative row-column approach, it is feasible to perform the training task on HE-encrypted data in a SIMD manner, which will be described in the following subsection. 

\begin{figure}[t]
    \centering
    \begin{subfigure}[b]{0.48\textwidth}
        \centering
        \includegraphics[width=\linewidth]{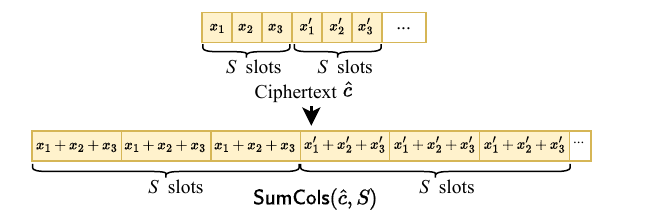}
        \caption{Illustration of $\textsf{SumCols}(.)$ algorithm. It first performs summation on all elements within each segment of the ciphertext. Then, the summation result is fit into each slot of the considered segment.}
        \label{fig:sumcol}
    \end{subfigure}%
    \hfill
    \begin{subfigure}[b]{0.46\textwidth}
        \centering
        \includegraphics[width=\linewidth]{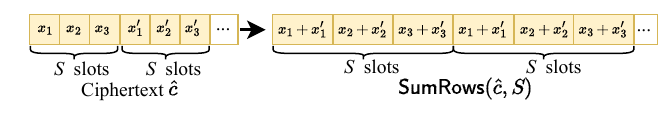}
        \caption{Illustration of $\textsf{SumRows}(.)$ algorithm. It first performs element-wise summation from different segments. This results in new segments with the same data where each slot of the segments is the combined sum of element $x_i$ from the original segments.}
        \label{fig:sumrow}
    \end{subfigure}
    \caption{Illustration of $\textsf{SumCols}(.)$ and $\textsf{SumRows}(.)$ algorithms on a ciphertext with multiple segments, each having $S$ slots size where each slot represents a distinct element $x_i$~\cite{lr2018rowcol}.}
    \label{fig:colrow}
\end{figure}

\subsubsection{The Proposed Training Process}

\begin{figure*}[t]
    \centering
    \includegraphics[width=0.9\linewidth]{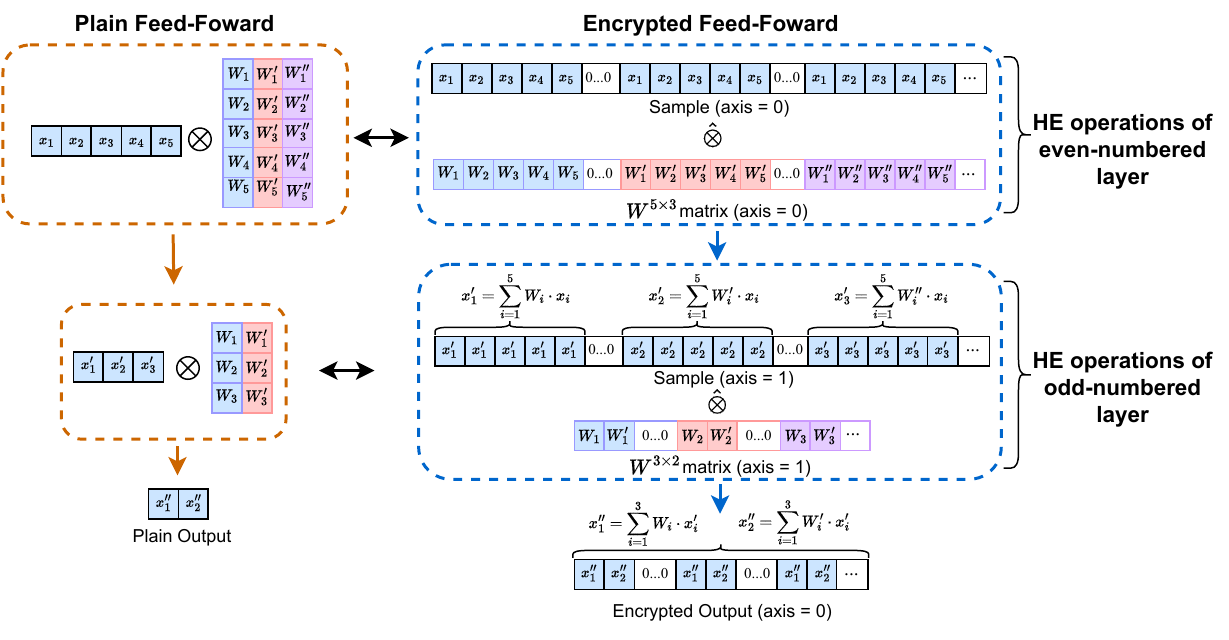}
    \caption{Implementation of encrypted feed-forward in a neural network with two layers. The weight matrix of the neural network layer is packed and encrypted based on the $k$-th order of the considered layer, regarding normal processing (axis=0) or transpose processing (axis=1). Therefore, through the HE multiplication described in~(\ref{eq: mul}), the input ciphertext, initially packed along axis=0, can alternatively fit into the encrypted neural network layers.}
    \label{fig:foward}
\end{figure*}

During the training task, while performing the homomorphic evaluation of the encrypted parameters, we utilize the \textsf{SumCols} and \textsf{SumRows} algorithms in~\cite{lr2018rowcol}. These algorithms take input as an encrypted vector and $S$ slot size to allow the sum operation of multiple segments on a different axis (i.e., rows or columns) of the encrypted vector, which is further described in Fig~\ref{fig:colrow}. After finishing generating the encrypted parameters of the neural network, the training process starts by initially forwarding the encrypted data through the encrypted parameters. 
% We denote the mini-batch of encrypted samples from $\hat{\mathcal{D}}$ as $\mathbb{B}$ which $\mathbb{B} = \{ (\bm{\hat{x}}_1^v, \bm{\hat{y}}_1^v),\dots, (\bm{\hat{x}}_m^v, \bm{\hat{y}}_m^v)\}$ with $m < I$, and $v=\left\lceil \frac{I}{m} \right\rceil$. 
In the feed-forward process, the output at layer $k$ of the neural network, denoted as $\bm{\hat{h}}^{(k)}$, can be calculated as:

\begin{equation}
    \bm{\hat{h}}^{(k)} = \hat{\sigma}^{(k)}\left(\hat{\bm{W}}^{(k)} \:\hat{\otimes}\: \bm{\hat{h}}^{(k-1)} \:\hat{\oplus}\: \hat{\bm{b}}^{(k)}\right),
\end{equation}
where $\hat{\bm{W}}^{(k)}$ and $\hat{\bm{b}}^{(k)}$ represent the encrypted weight matrix and encrypted bias vector of layer $k$, as explained in~(\ref{eq:enc_weight}) and~(\ref{eq:enc_bias}), respectively. $\hat{\otimes}$ and $\hat{\oplus}$ are, respectively, the homomorphic multiplication and addition for encrypted parameters of the neural network, which can be defined as:
\begin{equation}
\label{eq: mul}
    \hat{\bm{W}} \:\hat{\otimes}\: \hat{\bm{x}} = 
    \begin{cases}
        \textsf{SumCols}\big( \textsf{Mult}(\hat{\bm{W}}, \hat{\bm{x}}), S\big), & \text{if } \hat{\bm{W}}_{\text{axis}} = 0\\
        \textsf{SumRows}\big( \textsf{Mult}(\hat{\bm{W}}, \hat{\bm{x}}), S\big), & \text{if } \hat{\bm{W}}_{\text{axis}} = 1
    \end{cases}
\end{equation}

\begin{equation}
\label{eq: add}
    \hat{\bm{W}} \:\hat{\oplus}\: \hat{\bm{x}} = \textsf{Add}(\hat{\bm{W}}, \hat{\bm{b}}),
\end{equation}
where $\hat{\oplus}$ and $\hat{\otimes}$ are basically based on the \textsf{Add}(.), \textsf{Mult}(.), \textsf{SumCols}(.), and \textsf{SumRows} algorithms. To be more specific, after multiplying the input ciphertext $\hat{\bm{x}}$ with $\hat{\bm{W}}$, the $\hat{\otimes}$ proceeds summation based on the axis of $\hat{\bm{W}}$, which is illustrated in~(\ref{eq: mul}). The sum ciphertext is then applied \textsf{Add}(.) with $\hat{\bm{b}}$ to form the final output ciphertext. Subsequently, this ciphertext is passed through $\hat{\sigma}$, which illustrates the polynomial approximation of the activation function. As HE only supports homomorphic evaluations (e.g., addition, multiplication), the conventional activation function, such as \textit{ReLU} or \textit{Tanh}, may not be efficient for HE operations~\cite{lee2022cheb}. Accordingly, in this paper, we employ the Chebyshev polynomial method~\cite{lee2022cheb} to approximate the \textit{Swish} (SiLU) function~\cite{elwing2018silu}. The SiLU is chosen for its advantage in mitigating the ``dying \textit{ReLU}" problem. To be more specific, the general forward process of our proposed training algorithm is described in Fig.~\ref{fig:foward}. Particularly, input 1D data is encoded using Pack1D, while weight matrices are encoded using Pack2D, alternating between axes: even-numbered layers along the horizontal axis and odd-numbered layers along the vertical axis. Although the biases are not shown in the figure for simplicity, they are encoded on the opposite axis of their weight matrices. By leveraging the equations~(\ref{eq: mul}) and~(\ref{eq: add}), horizontally encoded input ciphertexts from even-numbered layers are forwarded into vertically encoded outputs, ensuring compatibility with the subsequent odd-numbered layer.

In the backpropagation process, we leverage the Stochastic Gradient Descent (SGD) for mini-batch and the Mean Squared Error (MSE) loss to enable the back-propagation of the encrypted data. Given a batch with $B$ pairs of predicted label $\hat{\bm{a}}_i$ and true label $\hat{\bm{y}}_i$, the MSE loss for optimizing encrypted model parameters during training can be defined as:
\begin{equation}
    \hat{\mathcal{L}}_{MSE} = \textsf{Mult}\Big(\sum_{i=1}^B \textsf{Square}\big(\textsf{Sub}(\bm{\hat{y}}_i, \bm{\hat{a}}_i) \big), \frac{1}{B} \Big),
\end{equation}
in which $\textsf{Square}(\bm{\hat{c}}) = \textsf{Mult}(\bm{\hat{c}},\bm{\hat{c}})$. Following that, the gradient of the encrypted MSE loss can be denoted as:
\begin{equation}
\label{eq:loss_grad}
    \nabla \hat{\mathcal{L}}_{MSE,i} = \frac{\partial \hat{\mathcal{L}}_{MSE}}{\partial \bm{\hat{a}}_i} = \textsf{Mult}\Big(\textsf{Sub}(\bm{\hat{y}}_i, \bm{\hat{a}}_i), \frac{2}{B} \Big).
\end{equation}
The calculated gradient $\nabla \hat{\mathcal{L}}_{MSE,i}$ is then utilized to calculate the gradients for $k$ encrypted layers. Given that $K$ is the latest layer of the neural network, the backward process starts by initially calculating the encrypted gradient of layer $K$:

\begin{equation}
\label{eq:grad}
    \nabla \hat{\bm{g}}_i^{(K)} = \textsf{Mult} \Big(\nabla \hat{\mathcal{L}}_{MSE,i}, \frac{\partial \hat{\sigma}^{(K)}}{\partial \bm{\hat{h}}^{(K-1)}_i}\Big).
\end{equation}
% in which $\bm{\hat{x}}_{\text{out},i}$ is the encrypted output of layer $K$ before fitting into the activation function.

\begin{algorithm}[t]
\caption{Iterative Training Algorithm for HE-Encrypted Data}
\label{ago:training}
\begin{algorithmic}[1]
\State \textbf{Input:} Initialize the encrypted data $\hat{\mathcal{D}}=(\bm{\hat{x}}_i, \bm{\hat{y}}_i)$ based on~(\ref{eq:enc_x}),~(\ref{eq:enc_y}); $i \in I$
\State Initialize the encrypted parameters $\hat{\theta}=(\bm{\hat{W}}^{(k)}, \bm{\hat{b}}^{(k)})$  based on~(\ref{eq:enc_weight}),~(\ref{eq:enc_bias}); $k \in K$
\State Initialize $T$ training rounds, learning rate $\eta$, and the HE parameters including (\textsf{pk},\textsf{sk}), $S$, $\mathcal{R}$
\For {$ t=0 \rightarrow T-1$}
    \State  Compute the output $\bm{\hat{a}}^{(k)}$ over layer $k$ using~(\ref{eq: mul}),~(\ref{eq: add}): 
    \begin{center}
        $\bm{\hat{a}}^{(k)}_t = \hat{\sigma}^{(k)}\left(\hat{\bm{W}}^{(k)}_t \:\hat{\otimes}\: \hat{x}^{(k-1)} \:\hat{\oplus}\: \hat{\bm{b}}^{(k)}_t\right)$
    \end{center}
    % \If{$\hat{\bm{W}}_{\text{axis}}^{(k)} = 0$}
    % \begin{center}
    %     $ \hat{\bm{W}}_t \:\hat{\otimes}\: \hat{x}_t =  \textsf{SumCols}\big( \textsf{Mult}(\hat{\bm{W}}_t, \hat{\bm{x}}_t), S\big)$
    % \end{center}
    % \Else 
    %     \begin{center}
    %     $ \hat{\bm{W}}_t \:\hat{\otimes}\: \hat{x}_t =  \textsf{SumRows}\big( \textsf{Mult}(\hat{\bm{W}}_t, \hat{\bm{x}}_t), S\big)$
    % \end{center}
    % \EndIf
     \State \parbox[t]{0.8\linewidth} {Compute gradient $\hat{\mathcal{L}}_{MSE,i}^t$ for all mini-batch of $B$ samples in $\hat{\mathcal{D}}$:}
     \begin{center}
        $ \nabla \hat{\mathcal{L}}_{MSE,i}^t = \textsf{Mult}\big(\textsf{Sub}(\bm{\hat{y}}_i, \bm{\hat{a}}_{i,t}), \frac{2}{B} \big)$
    \end{center}
    \State Compute gradient of layer $K$:
    \begin{center}
        $ \nabla \hat{\bm{g}}^{(K)}_{i,t} = \textsf{Mult} \Big(\nabla \hat{\mathcal{L}}_{MSE,i}^t, \frac{\partial \hat{\sigma}^{(K)}}{\partial \bm{\hat{x}}^{(K-1)}_i}\Big)$
    \end{center}
    \State Backward the gradient for $k$ layers:
    \begin{center}
        $\nabla \hat{\bm{g}}^{(k)}_{i,t} = \textsf{Mult}(\nabla \hat{\bm{g}}^{(k+1)}_{i,t}, \hat{\bm{W}}^{(k+1)}_{t})$
    \end{center}
    \State Produce final gradient $\nabla \hat{\bm{g}}^{(k)}_{i,t}$ using~(\ref{eq:backward_g})
    % \If{$\bar{\bm{W}}_{\text{axis}}^{(k+1)} = 0$}
    % \begin{center}
    %     $ \nabla \hat{\bm{g}}^{(k)}_i =  \textsf{SumRows}\big( \nabla \hat{\bm{g}}^{(k)}_i, S\big)$
    % \end{center}
    % \Else 
    %     \begin{center}
    %     $ \nabla \hat{\bm{g}}^{(k)}_i =  \textsf{SumCols}\big( \nabla \hat{\bm{g}}^{(k)}_i, S\big)$
    % \end{center}
    % \EndIf
    \State Compute:
    \begin{center}
        $\nabla \bm{\hat{W}}^{(k)}_t = \sum_{i=1}^B \textsf{Mult}(\bm{\hat{x}}_i^{(k-1)}, \nabla \hat{\bm{g}}_{i,t}^{(k)})$
    \end{center}
    \State and:
    % \State and $ \nabla \bm{\hat{b}}^{k}_t = \sum_{i=1}^B \nabla \hat{\bm{g}}^{(k)}_{i,t}$
    \begin{center}
        $ \nabla \bm{\hat{b}}^{k}_t = \sum_{i=1}^B \nabla \hat{\bm{g}}^{(k)}_{i,t}$
    \end{center}
    \State Update the $\hat{\theta}$ by SGD for encrypted data:
    \begin{center}
        $\bm{\hat{W}}^{(k)}_t \leftarrow\textsf{Sub}\Big(\bm{\hat{W}}^{(k)}_t,\textsf{Mult}\big(\eta, \nabla \bm{\hat{W}}^{(k)}_t\big)\Big)$
    \end{center}
    % \State and: $\bm{\hat{b}}^{(k)}_t \leftarrow \textsf{Sub}\Big(\bm{\hat{b}}^{(k)}_t,\textsf{Mult}\big(\eta, \nabla \bm{\hat{b}}^{(k)}_t\big)\Big)$
    \State and:
    \begin{center}
        $\bm{\hat{b}}^{(k)}_t \leftarrow \textsf{Sub}\Big(\bm{\hat{b}}^{(k)}_t,\textsf{Mult}\big(\eta, \nabla \bm{\hat{b}}^{(k)}_t\big)\Big)$
    \end{center}
    \State Bootstrap the encrypted weights and biases:
    \begin{center}
        $\bm{\hat{W}}^{(k)}_t = \textsf{Bootstrap}(\bm{\hat{W}}^{(k)}_t)$
    \end{center}
    \State and:
    % \State $\bm{\hat{b}}^{(k)}_t = \textsf{Bootstrap}(\bm{\hat{b}}^{(k)}_t)$
    \begin{center}
        $\bm{\hat{b}}^{(k)}_t = \textsf{Bootstrap}(\bm{\hat{b}}^{(k)}_t)$
    \end{center}
\EndFor
\end{algorithmic}
\end{algorithm}

Regarding the plain back-propagation, the gradient backward process requires the transpose matrix multiplication, which is challenging for the current HE-based neural networks. However, by employing our packing methods described in Algorithm~\ref{ago:pack1d} and Algorithm~\ref{ago:pack2d}, we can alternatively perform transpose multiplication between ciphertexts. Accordingly, 
the $\nabla \hat{\bm{g}}_i^{(K)}$ obtained from~(\ref{eq:grad}) is then backward to the previous $K-1$ layers. In particular, the backward of $\nabla \hat{\bm{g}}_i^{(K)}$ can be defined as:
\begin{equation}
    \nabla \hat{\bm{g}}_i^{(k)} = \textsf{Mult}(\nabla \hat{\bm{g}}_i^{(k+1)}, \hat{\bm{W}}^{(k+1)}),
\end{equation}
in which $\nabla \hat{\bm{g}}_i^{(k)}$ is then applied to the summation algorithms to produce the final gradient for layer $k$:
\begin{equation}
\label{eq:backward_g}
    \nabla \hat{\bm{g}}_i^{(k)} = 
    \begin{cases}
        \textsf{SumRows}\big( \nabla \hat{\bm{g}}_i^{(k)}, S\big), & \text{if } \bar{W}^{(k+1)}_{\text{axis}} = 0,\\
        \textsf{SumCols}\big( \nabla \hat{\bm{g}}_i^{(k)}, S\big), & \text{if } \bar{W}^{(k+1)}_{\text{axis}} = 1.
    \end{cases}
\end{equation} 
Subsequently, the $\nabla \hat{\bm{g}}^{(k)}$ is used to generate the $\nabla \bm{\hat{W}}^{(k)}$ and $\nabla \bm{\hat{b}}^{(k)}$ of layer $k$ which is calculated as:

\begin{equation}
    \nabla \bm{\hat{W}}^{(k)} = \sum_{i=1}^B \textsf{Mult}(\bm{\hat{h}}_i^{(k-1)}, \nabla \hat{\bm{g}}_i^{(k)}),
\end{equation}

\begin{equation}
    \nabla \bm{\hat{b}}^{(k)} = \sum_{i=1}^B \nabla \hat{\bm{g}}_i^{(k)},
\end{equation}
After generating the gradient for weight and bias, the SGD momentum is employed to update the parameters during the training process. The SGD update for encrypted parameters can be defined as:
\begin{equation}
    \bm{\hat{W}}^{(k)} \leftarrow \textsf{BootStrap}\Big(\textsf{Sub}\Big(\bm{\hat{W}}^{(k)},\textsf{Mult}\big(\eta, \nabla \bm{\hat{W}}^{(k)}\big)\Big)\Big),
\end{equation}

\begin{equation}
    \bm{\hat{b}}^{(k)} \leftarrow \textsf{BootStrap}\Big(\textsf{Sub}\Big(\bm{\hat{b}}^{(k)},\textsf{Mult}\big(\eta, \nabla \bm{\hat{b}}^{(k)}\big)\Big)\Big).
\end{equation}
in which the SGD optimizer for encrypted data uses the basic HE operations mentioned in~\ref{sec:he} with learning rate $\eta$ to update the weights and biases of the HE-based neural network iteratively. In addition, the bootstrapping mechanism is applied to the encrypted parameters to reduce the noise of ciphertext, allowing additional operations on encrypted data across multiple training rounds. To summarize, the proposed training algorithm for HE-encrypted data is described in Algorithm~\ref{ago:training}. This iterative training method will be employed in our proposed privacy-preserving distributed learning for cyberattack detection, which is described as follows.

\subsection{Implementation of Distributed Cloud-Native Learning}
\label{sec:implementation_ppdl}

As mentioned in Section~\ref{sec:pp_sys}, the BNs enter the offloading phase by initially generating the HE key pairs, which consist of the secret key $\textsf{sk}_n$ and the public key $\textsf{pk}_n$. The secret key is securely stored within the respective BNs, while the public key is broadcasted to the CSP and other nodes in the blockchain network. Besides validating transactions, BN-$n$ encrypt its local dataset $\mathcal{D}_n$ to $\hat{\mathcal{D}}_n$ by employing mechanisms in~(\ref{eq:enc_x})~(\ref{eq:enc_y}). Subsequently, the encrypted data is transmitted to the CSP, where it is combined to create a large encrypted dataset $\hat{\mathcal{D}}$. After that, the CSP generates the deep-learning model with parameters $\theta$ and encrypts it to $\hat{\theta}$ by using equations~(\ref{eq:enc_weight}) and~(\ref{eq:enc_bias}).

To start the encrypted data training phase, the CSP apply our proposed privacy-preserving distributed learning (PPDiL) by first determining the training parameters, including $T$ training rounds and batch size $B$. Concurrently, it leverages the MN and $M$ WNs within its cluster environment, which employs the computation from multiple workers while maintaining the MN as an aggregate point. Accordingly, the MN distributes the training parameters and encrypted model to each WN-$m$, with the local batch size $\left\lfloor \frac{B}{M} \right\rfloor$. The MN then divides the encrypted dataset $\hat{\mathcal{D}}$ into $M$ encrypted partitions $\hat{\mathcal{D}}_m$ and assigns each partition to respective WN-$m$. At training each round $t$, WN-$m$ employs the proposed training algorithm described in Algorithm~\ref{ago:training} to perform encrypted training on the corresponding $\hat{\mathcal{D}}_m$ and produce the encrypted trained model $\hat{\Theta}^t_m$. Afterwards, the workers transmit the $\hat{\Theta}^t_m$ to the MN, which then starts the aggregation by utilizing the FedAvg algorithm~\cite{mcmahan2017fedavg}, which is modified for encrypted data. In particular, it can be defined by:
\begin{equation}
    \hat{\Theta}_a^{(t+1)} = \textsf{Mult}\Big(\frac{1}{M},\sum_{m=1}^{M}\hat{\Theta}_{m}^t\Big) .
\end{equation}
At this point, the aggregated parameters $\hat{\Theta}_a^{(t+1)}$ are sent back to the WNs, which are subsequently updated to the workers' encrypted model for the next training round. This training process is iteratively maintained until the aggregated model converges and the CSP obtains the optimal parameters. 

\begin{algorithm}[t]
\caption{Privacy-Preserving Distributed Learning-Enabled Cyberattack Detection Framework}
\label{ago:ppdl_implement}
\begin{algorithmic}[1]
\For {$\forall n \in N$}
    \State \parbox[t]{0.8\linewidth}{$N$ Blockchain nodes generate the keypairs: $\textsf{sk}_n = \textsf{SKGen}(n)$ and $\textsf{pk}_n = \textsf{PKGen}(\textsf{sk}_n)$} 
    \State Extract $\mathcal{D}_n$ and process it by using Algorithm~\ref{ago:pack1d}
    \State Generate the encrypted data $\hat{\mathcal{D}}_n = \textsf{Enc}(\textsf{pk}_n, \mathcal{D}_n)$
    \State Send the encrypted data $\hat{\mathcal{D}}_n$ to the CSP
\EndFor
\State CSP combines received data into an encrypted dataset $\hat{\mathcal{D}}$
\State CSP initializes the deep learning model $\Theta$ and training parameters $T$, $B$
\State Pack $\Theta$ by using Algorithm~\ref{ago:pack2d} and generate the encrypted model $\hat{\Theta}$ where  $\hat{\Theta} = \textsf{Enc}(\textsf{pk}_n, \Theta)$
\State Distribute $\hat{\Theta}$, $B$, and $T$ to $M$ WNs
\State Split the encrypted dataset $\hat{\mathcal{D}}$ to $M$ partitions $\hat{\mathcal{D}}_m$ and assign each to respective WN-$m$
\While{$t \leq T$ or training process does not converge}
    \For{$\forall m \in M$}
    \State \parbox[t]{0.8\linewidth} {WN-$m$ calculates the trained parameters $\hat{\Theta}_m^t$ using Algorithm~\ref{ago:training}}
    \State \parbox[t]{0.8\linewidth} {WN-$m$ send their $\hat{\Theta}_m^t$ to the MN}
    \EndFor
\State MN produces the encrypted aggregated model $\hat{\Theta}_a^{(t+1)}$
\State Send the updated model $\hat{\Theta}_a^{(t+1)}$ back to $M$ WNs
\EndWhile
\State CSP transmits the optimized model $\hat{\Theta} = \hat{\Theta}^{(T)}$ to $N$ BNs
\For{$n \leq N$}
    \State Decrypt the model $\Theta = \textsf{Dec}(\hat{\Theta}, \textsf{sk}_n)$
    \State \parbox[t]{0.9\linewidth} {Predict incoming data based on the optimized model $\Theta$ in real-time}
    \vspace{0.8mm}
\EndFor
\end{algorithmic}
\end{algorithm}

After retrieving the optimized encrypted model $\hat{\Theta} = \hat{\Theta}^{(T)}$, the real-time cyberattack detection phase will commence. In this stage, the CSP transmit the encrypted model $\hat{\Theta}$ to the BNs for local deployment. After receiving $\hat{\Theta}$, BN-$n$ decrypt $\hat{\Theta}$ to $\Theta$ using its $\textsf{sk}_n$ and operate $\Theta$ as a cyberattack detection module to classify the incoming network traffic in real-time. Generally, the implementation of our proposed distributed learning algorithm is detailed in Algorithm~\ref{ago:ppdl_implement}.

% Given the blockchain nodes, including $F$ FNs and $L$ LNs, the CSP transmit the $\hat{\Theta}$ to the FNs for local deployment. After receiving $\hat{\Theta}$, FN-$f$ decrypt $\hat{\Theta}_f$ to $\Theta_f$ using its $\textsf{sk}_f$ and operate $\Theta_f$ locally to classify the incoming network traffic, denoted as $\mathcal{N}_{in}$. Regarding the LNs with limited computing resources, the network data is collected and transmitted to the CSP for detection tasks. Before sending data to the CSP, LN-$l$ encrypt $\mathcal{N}_{in,l}$ to $\hat{\mathcal{N}}_{in,l}$ by using the same packing method in the offline training phase. On the server side, CSP uses $\hat{\Theta}$ to make inference on $\hat{\mathcal{N}}_{in,l}$ to form the encrypted output $\hat{\mathcal{N}}_{out,l}$. The CSP sends back the output to LN-$l$, and LN-$l$ can view the result by decrypting it. Generally, the implementation of our proposed distributed learning algorithm is detailed in Algorithm~\ref{ago:ppdl_implement}.

\section{Performance Evaluation}
\label{sec: performance_eval}

\subsection{Simulation Setup and Evaluation Metrics}

\subsubsection{Dataset and PPDiL Parameters} To evaluate our proposed framework, we use the Blockchain network traffic data from the BNAT dataset~\cite{khoa2024bnat}, which is designed for cyberattack detection in blockchain-based IoT networks. The dataset is collected via the IoT gateways and the blockchain nodes, with IoT gateways transmitting IoT transactions to the blockchain nodes, thereby making it ideal for our considered system. The BNAT dataset contains normal traffic and four types of Blockchain network attacks: 
% \color{blue}
\begin{itemize}
    \item  Denial of Services (DoS): is a common type of attack where the attacker can launch massive traffic to a targetted blockchain node, overwhelming the overall blockchain network.
    \item Brute-force Password (BP): is a traditional attack approach, in which the attacker performs brute-forcing to steal the blockchain users' accounts and illegally access users' digital wallets. 
    \item Flooding of Transactions (FoT): is a type of blockchain attack designed to disrupt the consensus mechanism by overwhelming the network with spam transactions that carry meaningless data. Similar to DoS, its goal is to overburden the mining process and hinder block generation, causing delays and inefficiencies for blockchain nodes.
    \item Man-in-the-Middle (MitM): is an attack where the attacker secretly stands between the legitimate blockchain nodes, thereby obtaining users' information and possibly intercepting the blockchain messages. 
\end{itemize}
% \color{black}

\begin{figure}[t]
    \centering
    \includegraphics[width=0.55\linewidth]{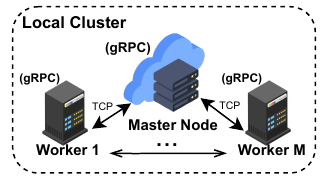}
    \caption{The experiment setup of CSP environment.}
    \label{fig:dl_experiment}
\end{figure}

% \begin{table}[t]
% \caption{Parameters setting}
% \label{tab: para}
% \centering
% \begin{tabular}{|c|>{\centering\arraybackslash}m{3cm}|>{\centering\arraybackslash}m{2cm}|}
% \hline
%  & \textbf{HE-Training Parameters} & \textbf{Values} \\ \hline
% $T$ & Training rounds & $T = 30$ \\ \hline
% $B$ & Initial batch size & $B = 128$ \\ \hline
% $Lr$ & Learning rate & $Lr = 0.9$ \\ \hline
% $\mathcal{R}$ & HE-Ring dimension & $\mathcal{R} = 10^{11}$ \\ \hline
% $\mathcal{B}$ & HE-ciphertext size & $\mathcal{B} = 1024$ \\ \hline
% $\mathcal{S}$ & HE-plain slot size & $S = 32$ \\ \hline

% \end{tabular}
% \end{table}

During the preprocessing, we downsampled the dataset into 10,000 samples, with 2,000 samples for each class. The dataset is then nominalized and min-max scaled within the range of (0,1). After that, we divide the dataset into training and testing sets with a ratio of 80:20, in which the training set is HE-encrypted, and the testing set is configured for two scenarios: non-encrypted inference and encrypted inference. Regarding the learning model configuration, we design a deep neural network containing an input layer, 2 hidden layers and an output layer. The number of neurons in the layers is 21, 32, 16, and 5, respectively. Apart from the input layer, each layer is connected to the polynomial approximation of the SiLU activation function. Following that, we utilize the OpenFHE library~\cite{openfhe} for HE and the Pytorch library for designing neural networks.

% \begin{figure}[t]
%     \centering
%     \includegraphics[width=0.55\linewidth]{Figures/DL_flwr.pdf}
%     \caption{The experiment setup of CSP environment.}
%     \label{fig:dl_experiment}
% \end{figure}

Due to the privacy-preserving distributed learning (PPDiL) setup, we select the Flower open-source~\cite{flower} for the experiment. We configure the CSP to have one master node and a maximum of five WNs operating by six different workstations within a local cluster environment. Three of these workstations are equipped with Intel Xeon E-2288G @3.7GHz processors with 8 cores. The remaining three workstations are powered by Intel Xeon Gold 6238R @2.2GHz processors with 28 cores (26 cores enabled). As described in Fig.~\ref{fig:dl_experiment}, the MN and WNs are deployed by the Google Remote Procedure Call (gRPC) server-client model, and their connections are maintained by TCP protocol with a connection timeout set at 4600 seconds. This cluster environment is employed to apply our proposed PPDiL for HE-encrypted data with the parameters illustrated in Table~\ref{tab: para}.

During the simulation, we evaluate our proposed distributed learning framework with different numbers of workers (i.e., two workers to five workers) and a centralized approach regarding processing time. Similar to~\cite{miran2018logistic}, we consider the training with non-encrypted data as the benchmark to analyze the effectiveness of the detection model after applying the proposed PPDiL. 

\begin{table}[t]
\caption{Parameters setting}
\label{tab: para}
\centering
\begin{tabular}{|>{\centering\arraybackslash}m{3cm}|>{\centering\arraybackslash}m{2cm}|}
\hline
\textbf{HE-Training Parameters} & \textbf{Values} \\ \hline
Training rounds & $T = 30$ \\ \hline
Initial batch size & $B = 128$ \\ \hline
Learning rate & $\eta = 0.9$ \\ \hline
HE-Ring dimension & $\mathcal{R} = 2^{11}$ \\ \hline
HE-ciphertext size & $\mathcal{B} = 2^{10}$ \\ \hline
HE-plain slot size & $S = 32$ \\ \hline
\end{tabular}
\end{table}

\subsubsection{Evaluation Metrics} To evaluate the performance of the detection model, the confusion matrix is utilized, which is suitable for a machine learning-based classification system~\cite{fawcett2006roc}. We denote $\mbox{TP}$, $\mbox{TN}$, $\mbox{FP}$, and $\mbox{FN}$ as, respectively, ``True Positive'', ``True Negative'', ``False Positive'', and ``False Negative''. Assuming the system consists of $C$ classes, which include normal and attack traffic, the accuracy can be calculated as:
\begin{equation} 
    \text{$Accuracy$} = \frac{1}{C} \sum_{c=1}^C \frac{\mbox{TP}_c + \mbox{TN}_c}{\mbox{TP}_c + \mbox{TN}_c + \mbox{FP}_c + \mbox{FN}_c}.
\end{equation}

The macro-average precision and recall are utilized in this term. The macro-average precision is:
\begin{equation}
    \text{$Precision$} = \frac{1}{C} \sum_{c=1}^C \frac{\mbox{TP}_c}{\mbox{TP}_c + \mbox{FP}_c}.
\end{equation}

The macro-average recall is calculated as follows:
\begin{equation}
    \text{$Recall$} = \frac{1}{C} \sum_{c=1}^C \frac{\mbox{TP}_c}{\mbox{TP}_c + \mbox{FN}_c}.
\end{equation}

\subsection{Simulation Results}
\begin{figure}[t]
    \centering
    \includegraphics[width=\linewidth]{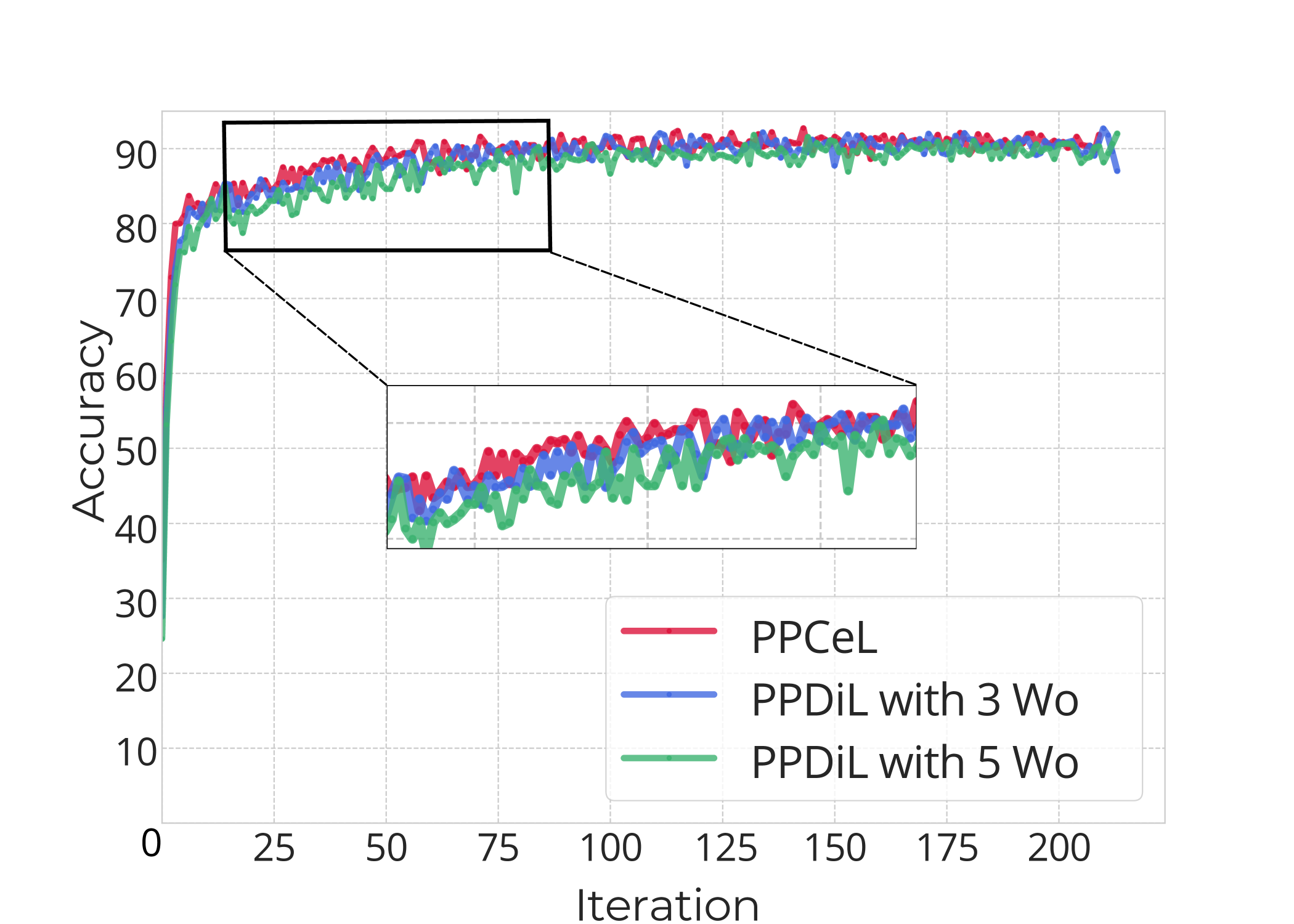}
    \caption{Convergence of the considered learning algorithms.}
    \label{fig:conv}
\end{figure}
\subsubsection{Convergence Analysis} In this subsection, we examine the convergence rate of our proposed PPDiL. Fig.~\ref{fig:conv} demonstrates the HE-encrypted training process of the centralized approach (PPCeL) and our proposed distributed approach (PPDiL), where the scenario of three workers and five workers are considered.
Accordingly, the PPCeL reaches convergence after around 75 iterations due to the processing of a large dataset at a centralized point. In contrast, the PPDiL exhibits a slower convergence rate than PPCeL, with the convergence speed decreasing as more workers join the process. In detail, the PPDiL with five workers requires nearly 90 iterations to achieve stable accuracy, while the PPDiL with three workers is nearly equivalent to the PPCeL, reaching convergence after 75 iterations. However, despite the gap in convergence speed, the accuracy of PPDiL and PPCeL remains consistent, regardless of whether the number of workers increases. To be more specific, both the accuracy of PPCeL and PPDiL remains the same after 200 iterations, stabilizing at around 91\%. As a result, moving from PPCeL to PPDiL training can still achieve effective accuracy during the learning process.

\subsubsection{Training Time Evaluation} As described in Fig.~\ref{fig:time}, we comprehensively analyze the improvement of our proposed PPDiL compared to the CeL approach. In Fig.~\ref{fig:totaltime}, we first provide the total training time of CeL and PPDiL in which distributing the training tasks to multiple workers shows an improvement in processing time. While the CeL requires 52.75 hours to finish the training, the proposed PPDiL can decrease the computation time significantly to 33.97, 27.11, 23.81, and 21.64 hours by distributing the learning tasks to two workers, three workers, four workers, and five workers, respectively.  
However, distributing the learning tasks among various workers introduces communication overhead, which we defined as the waiting time experienced by worker nodes after completing each learning round due to different workstation implementations. It is worth noting that this difference is based on the heterogenous of workstations, including differences in hardware capabilities, operating systems, software configurations, and varying conditions at different times.
As can be seen from Fig.~\ref{fig:totaltime}, the PPDiL with two workers has low communication overhead because the PPDiL only need to maintain the connection between a given master node and two WNs. Meanwhile, scenarios involving more workers incur larger communication overhead due to the varying hardware capabilities of the different participants. This discrepancy results in faster workers having to wait for slower ones to complete their rounds, leading to delays in the synchronization time during the training phase. 
% Following that, Fig.~\ref{fig:commover} provides a more detailed analysis of the communication overhead within one training round. Specifically, the aggregation process of PPDiL remains nearly identical regardless of the number of workers, maintained around 15 seconds for each round. Therefore, the disparity mostly comes from the synchronization time. As shown in Fig.~\ref{fig:commover}, with two workers in the PPDiL setup, the average waiting time for model transmission and aggregation is only 57.5 seconds. In contrast, the other counterparts (i.e., PPDiL with 3 to 5 workers) require from 121.45 seconds to 188.57 seconds for the workers to be synchronized, approximately 2 to 3 minutes per round. 

\begin{figure}[t]
    \centering
    \begin{subfigure}[b]{0.4\textwidth}
        \centering
        \includegraphics[width=\linewidth]{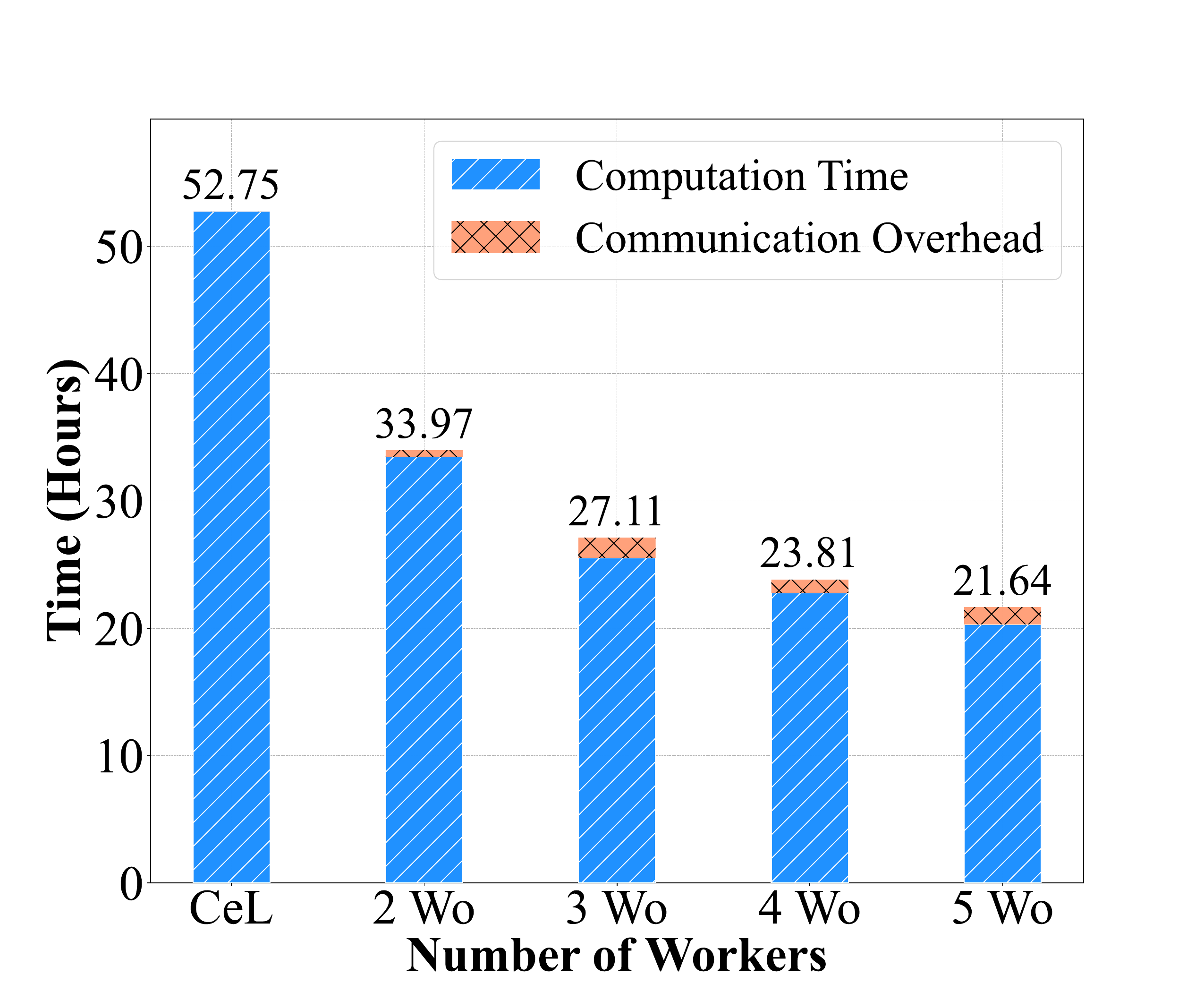}
        \caption{Total runtime during training process}
        \label{fig:totaltime}
    \end{subfigure}%
    \hfill
    % \begin{subfigure}[b]{0.33\textwidth}
    %     \centering
    %     \includegraphics[width=\linewidth]{Figures/Comm_overhead.pdf}
    %     \caption{Communication overhead in one round}
    %     \label{fig:commover}
    % \end{subfigure}
    \begin{subfigure}[b]{0.4\textwidth}
        \centering
        \includegraphics[width=\linewidth]{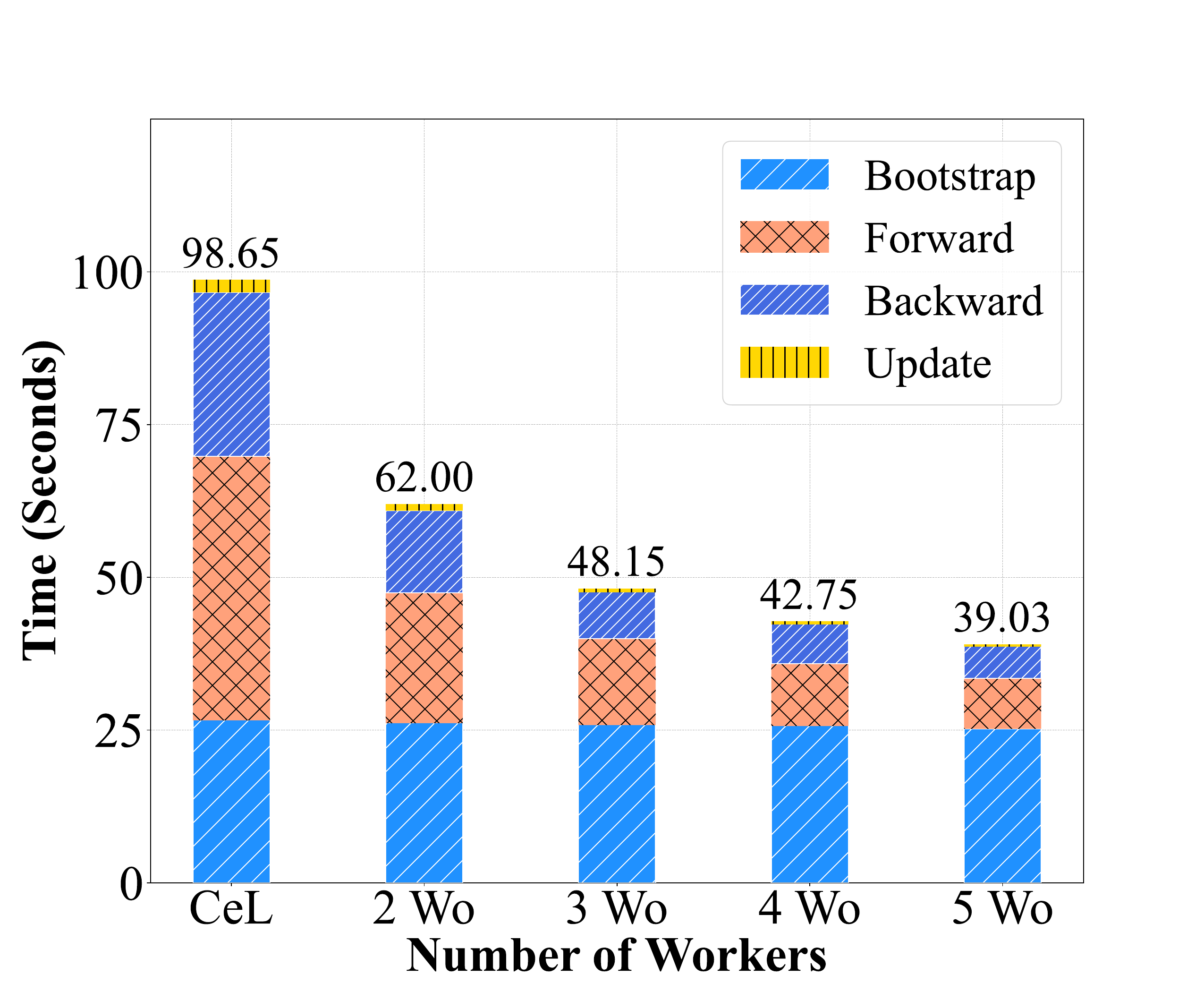}
        \caption{Training time in one iteration}
        \label{fig:itertime}
    \end{subfigure}%
    \hfill
    \caption{Execution time of offline training phase.}
    \label{fig:time}
\end{figure}

Additionally, Fig.~\ref{fig:itertime} provide a comprehensive analysis of the computation time over different stages during training (i.e., forward, backward, update, and bootstrap) in one training iteration. As the bootstrap mechanism is only applied to the neural networks' layers, its processing time remains consistent across CeL and other PPDiL scenarios, averaging around 25 seconds. For other training tasks, the PPDiL demonstrates a significant advantage by linearly reducing computation time, from 98.65 seconds with CeL down to 39.03 seconds with 5 workers. It is worth noting that aside from the bootstrap time, the processing time for other tasks in PPDiL decreases exponentially as the number of distributed workers increases, particularly in comparison to CeL.

To be more specific, Table~\ref{tab: time_iter} illustrates a more detailed analysis of Fig.~\ref{fig:itertime} regarding computational over each layer. 
% Generally, the computation over the main components of the deep learning model, including layers and activation functions, decreases exponentially as the number of distributed workers increases. 
Considering the forward process of CeL and PPDiL with 5 workers, the computation times for CeL over the input layer, hidden layer, and output layer are 10.376 seconds, 5.145 seconds, and 8.074 seconds, respectively. In contrast, the forward process time for PPDiL with 5 workers over the corresponding neural network layers is significantly reduced to 1.923 seconds, 0.936 seconds, and 1.485 seconds, which represent one-fifth of the computation times for CeL. However, the bootstrapping during the update process remains identical to both CeL and DL (i.e., with 2 to 5 workers) approaches, maintained from around 8 to 9 seconds. Therefore, this leads to a non-exponential decrease in training time. As a result, based on our real implementation of the proposed PPDiL framework, despite the impact of the bootstrapping time and the communication overhead, the training time is still improved. The participation of additional workers in the training process leads to significantly faster learning time compared to the centralized learning approach.

\begin{table}[t]
\centering
\caption{Processing time over one training iteration}
\label{tab: time_iter}
\begin{tabular}{|p{0.8cm}|c|c|c|c|c|c|}
\hline
\multirow{2}{*}{ \textbf{\makecell{Pro- \\ cess}}} & \multirow{2}{*}{\textbf{Layer}} & \multicolumn{5}{c|}{\textbf{Time (seconds)}} \\ \cline{3-7}
 & & \textbf{CeL} & \textbf{2 Wo} & \textbf{3 Wo} & \textbf{4 Wo} & \textbf{5 Wo} \\ \hline
\multirow{6}{*}{\centering \makecell{For- \\ ward}} & Input Layer &10.376 &5.140 &3.394 &2.416 &1.923 \\ \cline{2-7}
 & SiLU 1 &6.954 &3.487 &2.223 &1.698 &1.455 \\ \cline{2-7}
 & Hidden Layer &5.145 &2.478 &1.643 &1.174 &0.936 \\ \cline{2-7}
 & SiLU 2 &8.451 &4.222 &2.752 &1.965 &1.561 \\ \cline{2-7}
 & Output Layer &8.074 &3.885 &2.695 &1.852 &1.485 \\ \cline{2-7}
 & SiLU 3 &4.164 &2.102 &1.357 &1.062 &0.867 \\ \hline
\multirow{3}{*}{\makecell{Back- \\ ward}} & Output Layer &6.993 &3.521 &2.228 &1.726 &1.384 \\ \cline{2-7}
 & Hidden Layer &12.325 &6.209 &4.060 &2.845 &2.329 \\ \cline{2-7}
 & Input Layer &7.497 &3.713 &2.417 &1.876 &1.526 \\ \hline
\multirow{6}{*}{Update} & Input Layer &0.695 &0.324 &0.171 &0.116 &0.095 \\ \cline{2-7}
 & Bootstrap 1 &8.874 &8.591 &8.623 &8.547 &8.472 \\ \cline{2-7}
 & Hidden Layer &0.727 &0.394 &0.201 &0.150 &0.125 \\ \cline{2-7}
 & Bootstrap 2 &8.778 &8.713 &8.619 &8.676 &8.429 \\ \cline{2-7}
 & Output Layer &0.589 &0.341 &0.189 &0.145 &0.119 \\ \cline{2-7}
 & Bootstrap 3 &9.008 &8.877 &8.637 &8.502 &8.327 \\ \hline
\end{tabular}
\end{table}

\begin{table*}[htbp]
\centering
\caption{Performance comparisons of CeL/DiL and PPCeL/PPDiL with non-encrypted detection}
\label{tab:training_simu}
\begin{tabulary}{\textwidth}{|C|C|C|C|C|C|C|C|C|C|C|}
\hline
\multirow{3}{*}{\textbf{Model}} & \multicolumn{2}{c|}{\textbf{Centralized}} & \multicolumn{8}{c|}{\textbf{Distributed}} \\ \cline{2-11}
                                & \multirow{2}{*}{\textbf{CeL}} & \multirow{2}{*}{\textbf{PPCeL}} & \multicolumn{2}{c|}{\textbf{2 Wo}} & \multicolumn{2}{c|}{\textbf{3 Wo}} & \multicolumn{2}{c|}{\textbf{4 Wo}} & \multicolumn{2}{c|}{\textbf{5 Wo}} \\ \cline{4-11}
                                & & & \textbf{DiL} & \textbf{PPDiL} & \textbf{DiL} & \textbf{PPDiL} & \textbf{DiL} & \textbf{PPDiL} & \textbf{DiL} & \textbf{PPDiL} \\ \hline
\textbf{Accuracy}    & 91.487 & \textbf{91.450} & 91.483 & \textbf{91.375} & 91.475 & \textbf{91.350} & 91.016 & \textbf{90.875} & 91.024 & \textbf{90.925} \\ \hline
\textbf{Precision}   & 91.796 & \textbf{91.749} & 91.921 & \textbf{92.092} & 92.071 & \textbf{91.610} & 91.505 & \textbf{91.343} & 91.431 & \textbf{91.433} \\ \hline
\textbf{Recall}      & 91.507 & \textbf{91.465} & 91.441 & \textbf{91.339} & 91.476 & \textbf{91.435} & 91.036 & \textbf{90.941} & 91.086 & \textbf{90.931} \\ \hline
\end{tabulary}
\end{table*}

\begin{table}[t]
\centering
\caption{Detection result of PPDiL with encrypted data}
\label{tab:performance_metrics}
\begin{tabulary}{\textwidth}{|C|C|C|C|C|C|C|C|C|C|}
\hline
\multirow{2}{*}{\textbf{Model}} & \textbf{PPCeL} & \multicolumn{4}{c|}{\textbf{PPDiL}} \\ \cline{3-6}
                                &                      & \textbf{2 Wo} & \textbf{3 Wo} & \textbf{4 Wo} & \textbf{5 Wo}\\ \hline
\textbf{Accuracy}    & 91.725               & 91.450         & 90.301         & 90.875         & 90.805  \\ \hline
\textbf{Precision}   & 92.002               & 92.293         & 91.550         & 91.369         & 91.337  \\ \hline
\textbf{Recall}      & 91.729               & 91.382         & 91.378        & 90.944         & 90.828  \\ \hline
\end{tabulary}
\end{table}

\subsubsection{Cyberattack Detection Ability Evaluation} In this subsection, we further demonstrate the accuracy of the proposed PPDiL with the classification results of non-encrypted and encrypted data. Table~\ref{tab:training_simu} compares the non-encrypted data detection performance of the normal training and the privacy-preserving training regarding both centralized and distributed approaches. In general, the accuracy, precision, and recall of the learning model trained with non-encrypted data (i.e., CeL and DiL) and the ones trained with encrypted data (i.e., PPCel and PPDiL) remain nearly identical. Due to the centralized approach,  PPCeL achieves an accuracy of 91.45\%, which is only marginally lower than CeL's 91.487\%, with a negligible difference of approximately 0.04\%. This slight gap underscores that both methods consistently deliver around 91.4\% accuracy, demonstrating the effectiveness of HE-encrypted training. 
A similar trend is observed regarding our proposed distributed approaches, where the results are consistent in different scenarios compared to non-encrypted training benchmarks. Notably, the results of PPDiL maintain the same accuracy level as DiL, with the gap nearly from 0.01 to 0.02\%. For instance, in terms of distributed learning with three workers, the DiL achieves an accuracy of 91.475\%, which is slightly higher than the accuracy of 91.35\% from PPDiL. As a result, this minor gap is acceptable, demonstrating the reliability of our proposed PPDiL. Additionally, the results from Table~\ref{tab:training_simu} also indicate the comparison between centralized and distributed approaches. We can observe that the PPDiL achieves an accuracy level nearly identical to PPCeL, in which the gap across different scenarios (i.e., two Wo to five Wo) ranges from approximately 0.1 to 0.5. However, the classification results in Fig.~\ref{fig:main_cr} show that the detection performance of PPDiL for each attack type with five workers is still accurate compared to the PPCeL. 

% Table~\ref{tab:training_simu} describes the cyberattack detection performance in terms of non-encrypted inference. In general, the accuracy, precision, and recall of the learning model trained with non-encrypted data (i.e., N-EncCel and N-EnDL) and the ones trained with encrypted data (EncCel and EncDL) remain nearly identical. Regarding the centralized approach, the EncCel has an accuracy of 91.45\%, which is slightly less than the N-EncCel with 91.487\%, resulting in a gap of approximately 0.04\%. However, this can be seen as a trivial gap; thus, both of their results maintain around 91.4\%. A similar trend is observed regarding our proposed distributed approaches, where the results are consistent in different scenarios compared to non-encrypted benchmarks. Specifically, the gap between N-EnDL and EncDL is approximately from 0.01 to 0.02\%. For instance, in terms of distributed learning with 3 workers, the accuracy of N-EncDL is 91.475\%, which is slightly higher than the accuracy of 91.35\% from EncDL. As a result, the gap between N-EnDL and EncDL is approximately from 0.01 to 0.02\%, which is acceptable. 

Moreover, Table~\ref{tab:performance_metrics} shows the detection performance of PPDiL over the encrypted data. As can be seen, the accuracy of the encrypted detection is nearly the same as its non-encrypted counterparts, remaining around 91\% in different scenarios. Hence, the trivial gap in the aforementioned results is due to the HE evaluation, as the noise is added during the homomorphic computation over encrypted data. Despite the added noise during HE computation, the detection results of our learning model still consistently maintain between 91\% and 92\%. Consequently, our proposed method enables accurate detection for both raw data and encrypted data, offering the versatility needed for real-world applications where encrypted input classification is required. 

\subsection{Blockchain Nodes-enabled HE Computational Evaluation}

\subsubsection{Experiment Setup} To evaluate the reliability of our proposed cyberattack detection framework, we perform the experiment on a blockchain node (BN) with multiple consensus mechanisms regarding the network data encryption process while mining the IoT data. In the experiment, we consider the different batch sizes of the local network data, including 50, 500, 1000, and 2000 samples. The implementation consists of a blockchain node and an IoT transaction issuer in which the configuration of the devices and software for the implementation is described as follows:
\begin{itemize}
    \item The BN operates on processor Intel® Core i9-13900K (24 cores, 32 threads). The IoT transaction issuer is run on the processor Intel® Core i7 with 16GB RAM. 
    \item The BN with PoW and PoA mechanisms is launched by Go-Ethereum v1.10.14 (\textit{Geth})~\cite{geth} - an official open-source implementation of the Ethereum protocol. 
    \item The BN with PoS mechanism is launched by \textit{Geth} v1.14.6 and \textit{Prysm} v5.0.3~\cite{prysm} - an official implementation of the PoS mechanism in Ethereum 2.0. It is noted that in the PoS node, \textit{Geth} operates as the chain execution and \textit{Prysm} provides the PoS as a third-party software. 
\end{itemize}

\begin{figure}[t]
    \centering
    \begin{subfigure}[b]{0.42\textwidth}
        \centering
        \includegraphics[width=\textwidth]{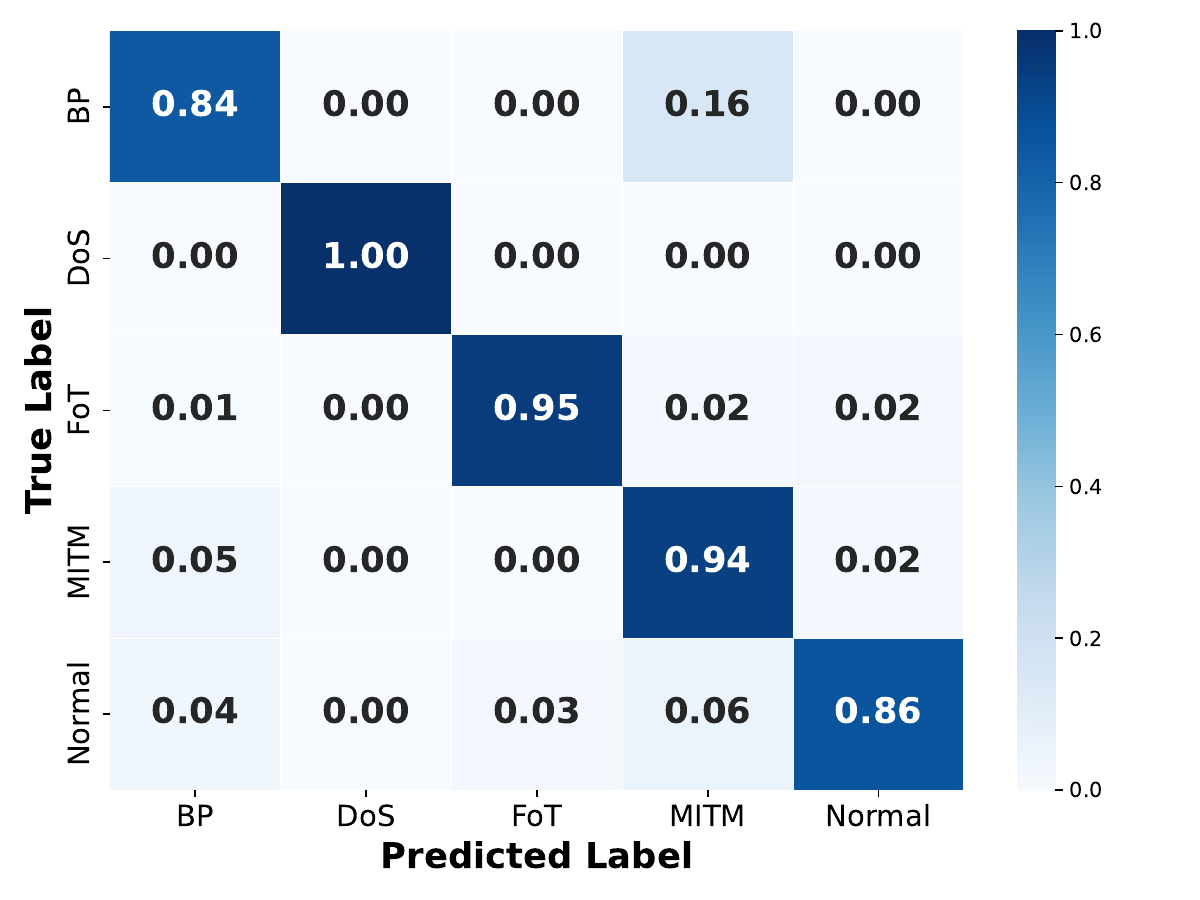}
        \caption{Confusion matrix of PPCeL}
        \label{fig:sub1}
    \end{subfigure}
    \hfill
    \begin{subfigure}[b]{0.42\textwidth}
        \centering
        \includegraphics[width=\textwidth]{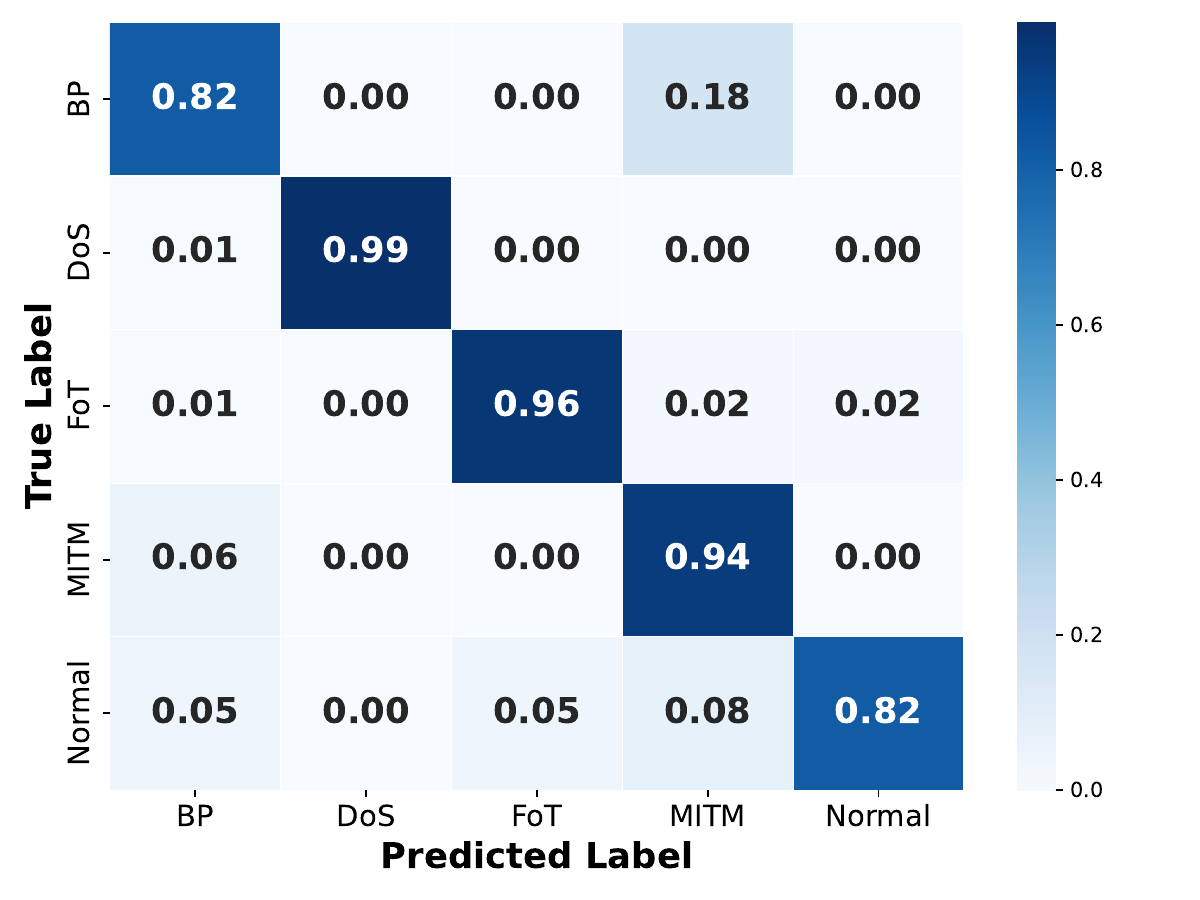}
        \caption{Confusion matrix of PPDiL with five workers}
        \label{fig:sub3}
    \end{subfigure}
    \caption{Classification results of the non-encrypted detection within the proposed HE-encrypted deep neural network.}
    \label{fig:main_cr}
\end{figure}

\subsubsection{Experiment Results} In Fig.~\ref{fig:node}, we examine our proposed framework with different mining algorithms in terms of computational resources and the latency time during the encryption process. As illustrated in Fig.~\ref{fig:resources}, the PoW blockchain node slightly costs higher resources during the encryption, with around 0.15 GB higher than other counterparts. Despite varying batch sizes, the encryption processes for PoA and PoS nodes consistently require computing resources nearly identical to those in scenarios without mining. Known for their lightweight nature, PoA and PoS maintain stable resource usage for encryption across 50, 500, 1000, and 2000 samples, holding at approximately 1.26 GB, 2.98 GB, 4.85 GB, and 8.75 GB, respectively. As a result, the mining process does not affect the resources of the BN during the HE-based encryption. 

To further demonstrate the scalability of our proposed framework, we analyze the latency time of the encryption during various scenarios. As observed in Fig.~\ref{fig:miningtime}, due to the intensive computational demands of PoW, the encryption process of the PoW-based BN incurs the highest latency, spanning from 5.28 seconds for 50 samples to 218.17 seconds for 2,000 samples. This results in latency that is approximately 8 to 12 times greater than that of a node operating without a mining process. In contrast to the trend in resources analysis, PoA and PoS incur slightly longer encryption times compared to nodes without a mining mechanism. In particular, the PoS node experiences delays ranging from 0.07 to nearly 7 seconds, primarily due to the reliance on third-party software to maintain the PoS algorithm, which reduces the resources available for data encryption. On the other hand, although PoA maintains lower latency than PoS, it still experiences slightly higher latency than the no-mining scenario, with minimal delays ranging from 0.06 to 1.69 seconds across different batch sizes. Consequently, while HE latency does increase across the three mining mechanisms, the slight latency difference in the lightweight algorithms (i.e., PoS and PoA) is minimal, suggesting that our proposed framework remains a viable option for various blockchain-based IoT networks.

\subsection{Encrypted Inference Time Analysis}

\begin{figure*}[t]
    \centering
    \begin{subfigure}[b]{0.48\textwidth}
        \centering
        \includegraphics[width=\textwidth]{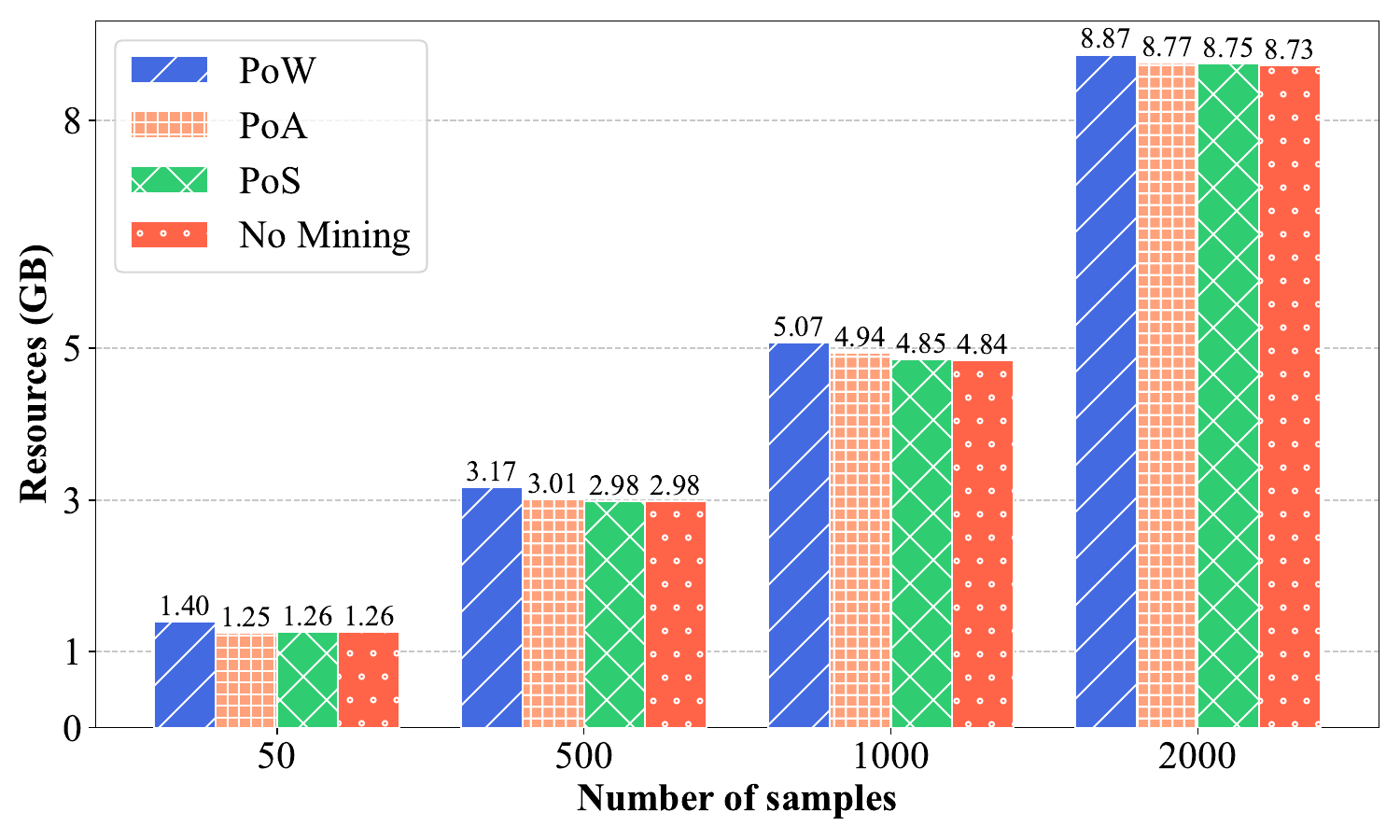}
        \caption{Computing resources of encryption with different batches size}
        \label{fig:resources}
    \end{subfigure}
    \hfill
    \begin{subfigure}[b]{0.48\textwidth}
        \centering
        \includegraphics[width=\textwidth]{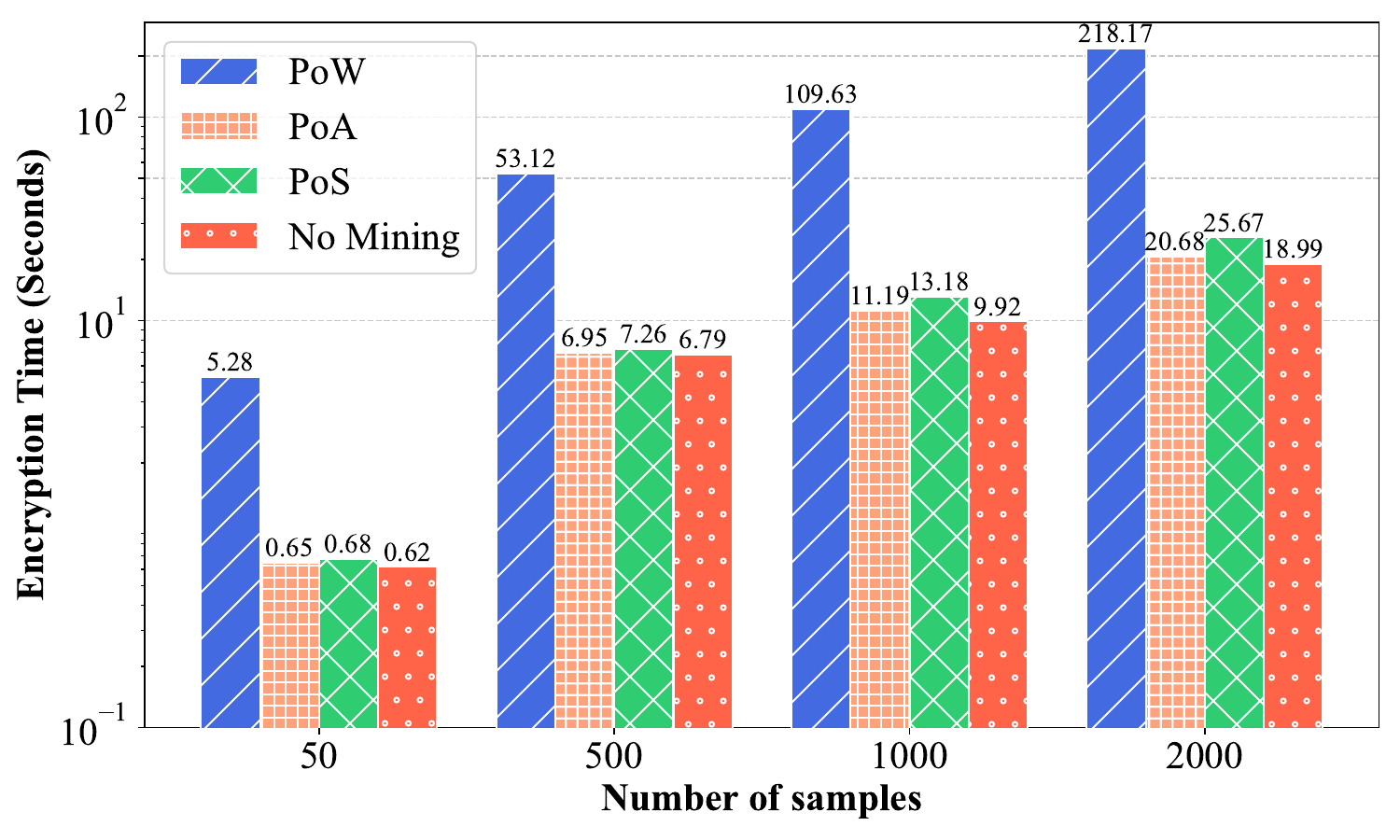}
        \caption{Latency of encryption within different batches size}
        \label{fig:miningtime}
    \end{subfigure}
    \caption{Evaluation of the \textit{Geth} blockchain node-enabled HE in different consensus mechanisms: PoW, PoA, PoS, and no mining.}
    \label{fig:node}
\end{figure*}
% \color{blue}
\subsubsection{Hardware Configuration} In this section, we perform the experiment to examine the adaptability of real-time detection with incoming encrypted data within our framework. During the implementation, we deploy the trained model, optimized through the proposed PPDiL, on the BN to detect the encrypted samples across various hardware configurations, which are described as follows:

\begin{itemize}
    \item BN-1: Intel Xeon E-2288G @3.7GHz processors with 8 processor cores.
    \item BN-2: Intel Xeon Gold 6238R @2.2GHz processors with 28cores (26 cores enabled).
    \item BN-3: AMD EPYC 9354P 3.25GHz processors with 32 cores 256MB L3 Cache.
\end{itemize}

Similar to the approach in~\cite{khoa2024bnat}, we consider the incoming network traffic data to be extracted in data frames, with each containing 400 samples for our experiment.

\begin{figure}[t]
    \centering
    \includegraphics[width=\linewidth]{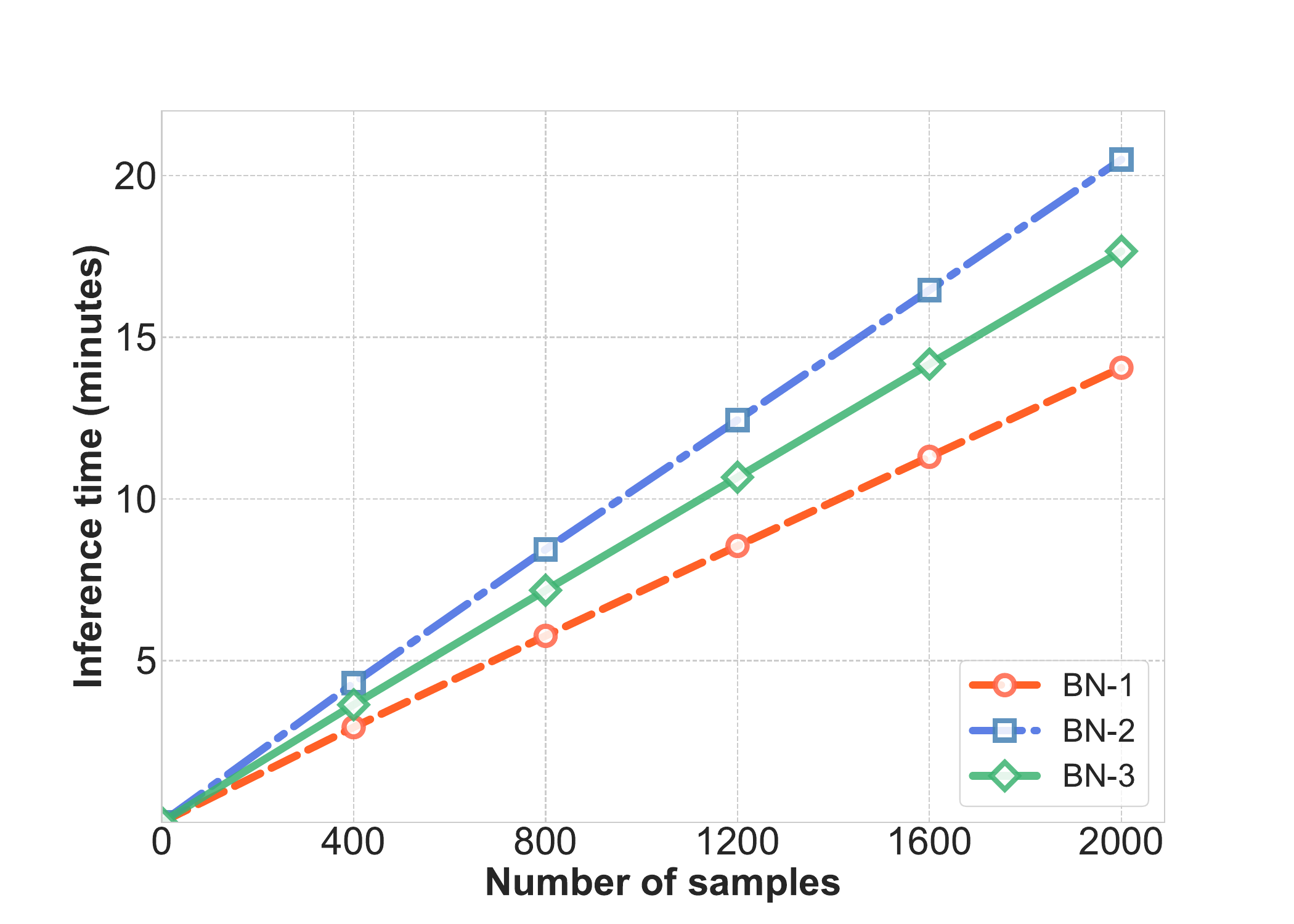}
    \caption{The inference time of the detection model with encrypted samples.}
    \label{fig:throughput}
\end{figure}

\subsubsection{Experiment Results} To analyze the effectiveness of our proposed cyberattack detection framework, we consider the processing time of the detection with encrypted samples. As observed in Fig.~\ref{fig:throughput}, the trained model employing our proposed PPDiL requires a maximum of nearly 5 minutes to predict the considered encrypted data frame with the deployment in BN-2 and BN-3. Meanwhile, regarding the utilization of BN-1 with more powerful hardware configurations, the inference time of the trained detection model shows significant improvement. Here, the trained model deployed on BN-1 can effectively detect the 400 encrypted samples in around 2.5 minutes, resulting in a throughput of approximately 2.4 samples per second. Therefore, the analysis indicates that our proposed cyberattack detection framework can be effectively adapted to real-world systems, especially with the enhanced computing power offered by modern edge servers\footnote{\url{https://aws.amazon.com/edge/}}.
% \color{black}

% \begin{table}[htbp]
% \centering
% \caption{Simulation results}
% \label{tab:time}
% \begin{tabulary}{\textwidth}{|C|C|C|C|C|C|C|}
% \hline
% \textbf{Model} & \textbf{Cel}   & \textbf{2w}  & \textbf{3w} & \textbf{4w}  & \textbf{5w}\\ \hline
% \textbf{Time (Hours)}    & 52.750  & 33.970  & 27.107  & 23.814  & 21.640\\ \hline
% \end{tabulary}
% \end{table}

% \begin{figure}[t]
%     \centering
%     \includegraphics[width=\linewidth]{Figures/Total_time2.pdf}
%     \caption{Execution time of offline training phase}
%     \label{fig:time}
% \end{figure}

% \begin{figure}[htbp]
%     \centering
%     \includegraphics[width=\linewidth]{Figures/Fl5wEnc_cm.pdf}
%     \caption{CM 5W}
%     \label{fig:cm5w}
% \end{figure}

\section{Conclusion}
\label{sec: conclu}

In this paper, we have proposed a novel privacy-preserving cyberattack detection framework for IoT-based blockchain networks, addressing computational challenges by allowing blockchain nodes to share IoT data with a cloud service provider (CSP) for training machine learning models. The framework leverages our proposed SIMD packing algorithms and CKKS encryption scheme to securely process and transmit data, enabling encrypted training on the CSP. We have introduced two complementary learning methods, including a deep neural network training algorithm and a distributed learning algorithm, both designed for efficiency and privacy. Our simulations and experiments have demonstrated that the proposed approach achieves 91\% accuracy with minimal processing time, closely matching the non-encrypted baseline, proving its potential for real-world application.

% or


% \appendices

% % use section* for acknowledgment
% \section*{Acknowledgment}

\ifCLASSOPTIONcaptionsoff
  \newpage
\fi



\bibliographystyle{IEEEtran}
\bibliography{library}

\end{document}